\documentclass[preprint,11pt]{article}

\usepackage{bm}
\usepackage{epsfig}
\usepackage{color}
\usepackage{amssymb}
\usepackage{amsmath}

\newcommand {\calleq}[1]{(\ref{eq:#1})}
\newcommand {\lab}[1]{\label{eq:#1}}

\newcommand {\be}[1]{\begin{equation}{\lab{#1}}}
\newcommand {\ee}{\end{equation}}
\newcommand {\bea}{\begin{eqnarray}}
\newcommand {\eea}{\end{eqnarray}}
\newcommand {\en}{\mathsf{e}}
\newcommand {\ov}[1]{\overline{#1}}

\begin{document}

\title{The two-stage dynamics in the Fermi-Pasta-Ulam \\ problem: 
from regular to diffusive behavior}

\author{
\textbf{  A. Ponno$^{a}$, H. Christodoulidi$^{a}$, Ch. Skokos$^{b}$, S. Flach$^{b}$ },\\
$^{a}$  Universit\`a degli Studi di Padova,\\
Dipartimento di Matematica Pura e Applicata, \\
Via Trieste 63, 35121 - Padova, Italy\\
$^{b}$ Max Planck Institut f\"{u}r Physik komplexer Systeme, \\
 N\"{o}thnitzer Str. 38, D-01187 Dresden, Germany}


\maketitle

\begin{abstract}
A numerical and analytical study of the relaxation to equilibrium of both the Fermi-Pasta-Ulam (FPU) $\alpha$-model and the integrable Toda model, when the fundamental mode is initially excited, is reported. We show that
the dynamics of both systems is almost identical on the short term, when the energies of the initially unexcited modes
grow in geometric progression with time, through a secular avalanche process. At the end of this first stage of the dynamics the time-averaged modal energy spectrum of the Toda system stabilizes to its final profile, well described, at low energy, by the spectrum of a $q$-breather. The Toda equilibrium state is clearly shown to describe well the long-living quasi-state of the FPU system. On the long term, the modal energy spectrum of the FPU system slowly detaches from the Toda one by a diffusive-like rising of the tail modes, and eventually reaches the equilibrium flat shape. We find a simple law describing the growth of tail modes, which enables us to estimate the time-scale to equipartition of the FPU system, even when, at small energies, it becomes unobservable.
\end{abstract}

                             
\newpage
\section{Introduction}

The present work deals with the so-called Fermi-Pasta-Ulam (FPU) problem, which consists in understanding and characterizing the process of dynamical relaxation to the micro-canonical equilibrium of non integrable Hamiltonian systems with many degrees of freedom, when initial conditions far from equilibrium are chosen. This issue, under investigation for more than 50 years, is relevant to equilibrium and out-of-equilibrium statistical mechanics, as well as to many problems in condensed matter physics.

Following the original work of FPU \cite{FPU}, one of the most studied systems has been, and presently is, the so-called 
$\alpha$-model, defined by the Hamiltonian
\be{H}
H_{\alpha}(q,p)=\sum_{n=0}^{N-1}\left[
\frac{p_n^2}{2}+\frac{(q_{n+1}-q_n)^2}{2}+
\alpha\frac{(q_{n+1}-q_n)^3}{3}
\right]\ ,
\ee
where $\alpha$ is a parameter and fixed ends
conditions are chosen: $q_0=q_N=p_0=p_N=0$.
Such a Hamiltonian model with $N-1$ degrees of freedom can be regarded either as
a space-discretized version of a nonlinear string 
(as FPU did), or as the expansion around the equilibrium configuration of a one-dimensional system of pairwise interacting identical particles,
and is commonly referred to as a nonlinear oscillator chain. 

FPU numerically integrated the equations of motion of 
system \calleq{H}, given by $\dot{q}_n=\partial H_{\alpha}/\partial p_n$,
$\dot{p}_n=-\partial H_{\alpha}/\partial q_n$,
and they chose initial conditions of the form
\be{incond}
q_n(0)=A\sin\left(\frac{\pi n}{N}\right)\ \ ,\ \ 
p_n(0)=0\ ,
\ee
$A$ being a parameter, thus initially exciting the fundamental (longest wavelength) Fourier mode of the chain. 
FPU considered system sizes $N=16,32,64$. The surprising and for a long time debated result they got, namely the absence of any complete
relaxation of the system to a state of equilibrium compatible with the laws of statistical mechanics (at least up to the observation times then available) was named, after them, \emph{the FPU paradox}. As first conjectured in Refs. \cite{Fetal,LPRVms} and then clearly shown in Ref. \cite{BGG}, 
the paradox is due to a separation of the relevant time-scales of the problem, which in turn depends on the value of the total energy $E$ and on the size $N$ of the system. More precisely, it has been understood that energy is first transferred to shorter wavelength 
modes, which, within a first characteristic time-scale $T^{qs}(E,N)$ gives rise to a \emph{quasi-state} (as named by FPU, later also referred to as ``metastable state'' \cite{Fetal,LPRVms} or ``natural packet'' \cite{BGG} in the literature). This state is characterized by a modal energy spectrum displaying an exponentially decreasing tail. On a second, larger time-scale 
$T^{eq}(E,N)$, the modal energy spectrum approaches an almost flat shape, i.e. the system does reach the equilibrium state characterized by modal energy equipartition, as predicted by the laws of equilibrium statistical mechanics. 
The FPU paradox appears then when, depending
on $E$ and $N$, a strong inequality $T^{qs}\ll T^{eq}$ holds, such that the approach to equilibrium becomes unobservable on the available computation times.

Up to now, the short term dynamics of the system has been thoroughly investigated, both from a numerical and an analytical point of view, also for initial conditions more general than \calleq{incond}, and both the time-scale $T^{qs}$ and the slope
of the exponential tail of the energy spectrum have been more or less quantified. In particular, the localization of the energy in Fourier space, i.e. the existence of the quasi-state, has been explained essentially in two ways. One approach consists in looking for a particular stable solution of the equations of motion that is close to the actual solution of the initial value problem and displays a localized energy spectrum.
One way to get such a solution consists in looking for the Lyapunov continuation of the initially excited mode, or $q$-breather \cite{FIK1,FIK2,FKIM,PF,FP}. 
The other approach consists instead in looking for a suitable integrable system whose dynamics is close, on a certain time-scale, to that of the actual system, and leads to the Korteweg-de Vries equation and its truncations in mode space
\cite{FP,ZK,Shep,P03,P05,BPcmp,BLP} (Ref. \cite{FP}
contains a first tentative comparison between the two approaches).
On the other hand, for the $\alpha$-model \calleq{H}, there exists an approximating integrable system given a priori, namely the Toda model (see Hamiltonian \calleq{HT} below), and a comparative study of the two models has been performed e.g. in Refs. 
\cite{FFML,ILRV,CCPC,GPP,Zab06}. 

Notwithstanding the results just quoted, both a systematic study of the dynamics of problem \calleq{H}-\calleq{incond} on short times and a deeper understanding of the possible existing links between the aforementioned theoretical approaches are still lacking. Such a gap is partially covered in the present paper. 
We explain the detailed resonance mechanisms ruling the energy cascade on short term together with a simple characterization of the quasi-state. Moreover, 
we compare, for the same initial conditions of the form \calleq{incond}, the dynamics of the FPU chain with that of the integrable Toda model\cite{TodaPR} defined by the Hamiltonian
\be{HT}
H_{T}(q,p)=\sum_{n=0}^{N-1}\left[
\frac{p_n^2}{2}+\frac{e^{2\alpha(q_{n+1}-q_n)}-1}{4\alpha^2}
\right]\ ,
\ee
with fixed ends at $n=0$ and $n=N$. The integrability of the latter Hamiltonian system was first suggested in Ref. \cite{FST} and then proved in Refs. \cite{He} and 
\cite{Fl}.
Expanding the exponential in \calleq{HT} and taking into account
the boundary conditions, one can write
\be{HaHT}
H_{\alpha}(q,p)=H_{T}(q,p)-\sum_{n=0}^{N-1}\sum_{r\geq4}
(2\alpha)^{r-2}\frac{(q_{n+1}-q_n)^r}{r!}\ ,
\ee
i.e. the FPU Hamiltonian can be regarded as a perturbation
of the Toda one. The call for such a comparative study for the evolution towards an equilibrium 
comes from the necessity to distinguish those phenomena that are of integrable - i.e. non chaotic - nature, from those that are instead signatures of non integrability - i.e. chaos. 
In the present paper we show that the FPU quasi-state coincides well with the Toda equilibrium state (see below, Sections II and III; see also Ref. \cite{FFML}), and is therefore a manifestation of closeness to integrability. Chaos becomes relevant in the $\alpha$-model only \emph{after} the saturation to the quasi-state. We also show that the quasi-state, i.e. the Toda equilibrium state, is well described, at low energy, by 
a time-periodic orbit or one-dimensional torus, namely
the $q$-breather associated to the first mode.

For what concerns the second time-scale $T^{eq}$, its numerical estimate has been given e.g. in Ref. \cite{CCPC} for the 
$\alpha$-model, and in Refs.
\cite{PL,DLR,BGP} for the so-called $\beta$-model; the two models are treated together and compared in the more recent Ref. 
\cite{BP11}, where accurate numerical estimates of $T^{eq}$ are given.
Yet, a detailed description of how the tail of the modal energy spectrum raises from zero to the equipartition level has not been provided so far. In the present paper, we perform a fit of the numerical data relative to the modal energy spectrum of the 
$\alpha$-model during the second part of its evolution, which yields a simple law describing the motion of the tail. 
This allows us
to measure the second time-scale $T^{eq}$ on times which can be much shorter than $T^{eq}$ (notice that
a direct measurement of the equipartition time-scale $T^{eq}$ becomes impossible when it  exceeds the available computation times).

We find that, at low energy (when the quasi-state is a one-dimensional Toda torus) the estimated time-scale $T^{eq}$ undergoes a transition from power law to stretched exponential of the inverse of the energy. As a consequence, the quasi-state of the $\alpha$-model becomes effectively stable: in practice the approach to equipartition becomes unobservable, which explains the FPU paradox.

Let us stress that the existing literature on the FPU problem is huge, encompassing more than fifty years of research on an unsolved problem. We thus defer the reader to some of the existing high-quality reviews on the subject\cite{FordPR,CHAOS,LNP}.

The paper is organized as follows. In Section II, after 
introducing some notations and a few relevant quantities,
we state our main results  on the evolution
towards equilibrium of systems 
\calleq{H} and \calleq{HT} with initial conditions \calleq{incond}.
In Section III, the phenomenology of the problem is illustrated by the numerical results and the fits with the analytical predictions.
In Section IV, the theory leading to the analytical estimates 
is reported. Finally, Section V is devoted to concluding remarks.

\section{Main results}

We integrated the equations of motion associated to the Hamiltonians \calleq{H} and \calleq{HT}
by means of a symplectic algorithm, namely the Yoshida kinetic-potential splitting algorithm of fourth order \cite{Yo1,Yo2},
with a time-step kept fixed to $0.05$ all over the explored energy range.\cite{N0}
The size $N$ of the system considered in the present paper is
kept fixed to $N=32$, and no systematic exploration of size
dependent phenomena is performed.
The chosen value of the nonlinearity parameter is 
$\alpha=0.33$.\cite{N1}

We follow the evolution of the harmonic energies $E_k$ of the Fourier modes of the system, also coined
\emph{modal energies}. These  quantities are defined as
\be{Ek}
E_k(t)\equiv\frac{P_k^2(t)+\omega_k^2Q_k^2(t)}{2}\ ,\ \ \ (k=1,\dots,N-1),
\ee
where $(Q_{k},P_{k})=
\sqrt{2/N}\sum_{n=1}^{N-1}(q_{n},p_{n})\sin(\pi k n/N)$
are the Fourier coordinates and 
\be{omk}
\omega_k=2\sin\left(\frac{\pi k}{2N}\right)
\ee
is the dispersion relation of the linearized problem 
($\alpha=0$). For the FPU model \calleq{H}, the total energy 
$E_{\alpha}=H_{\alpha}(q(0),p(0))$ in terms of the initial condition 
\calleq{incond} reads 
\be{E}
E_{\alpha}(A)=N\left(\frac{A\omega_1}{2}\right)^2\simeq 
\frac{\pi^{2}A^2}{4N}\ ,
\ee
whereas, for the same initial condition, the total energy 
$E_{T}=H_{T}(q(0),p(0))$ of the Toda model \calleq{HT} 
reads \cite{N2}
\be{ET}
E_{T}(A)=
\frac{N\left[I_{0}(2\alpha A\omega_{1})-1\right]}{4\alpha^{2}}\ ,
\ee
where $I_{0}(x)$ is the first modified Bessel function of order zero \cite{AbSt}. All the comparative runs were performed at the same value of the total energy $E$ for the two models, which required to consistently choose different initial amplitudes $A$ and $A'$ in the two cases: $E=E_{\alpha}(A)=E_{T}(A')$.
\cite{N3}

For $\alpha=0$ the modal energies $E_k$ do not evolve in time both for the FPU and the Toda model, since they are constants of motion: $E_1=E$ and $E_{s\geq2}=0$, for all times. When $\alpha>0$ the modal energies are no longer constants of motion for both systems, and evolve in time. 
Now, for the nonintegrable FPU system, one expects that each time-averaged modal energy
\be{Ekav}
\ov{E}_k(t)\equiv\frac{1}{t}\int_{0}^tE_k(s)\ ds
\ee
converges, as $t\rightarrow\infty$, to its final expectation value\cite{N4}, which is approximately equal to the specific energy 
$E/N$ of the system.

On the other hand, for the integrable Toda model the modal energies are quasi-periodic functions of time (with a maximum number $N-1$ of rationally independent frequencies), which ensures that their time-averages \calleq{Ekav} converge to some limit as $t\rightarrow\infty$. Such a limit
defines what we refer to as the \emph{equilibrium state of the Toda system}.  
In the sequel we use and plot the normalized modal energies
\be{ek}
\en_k(t)\equiv\frac{E_k(t)}{E}
\ee
and their time-averages $\overline{\en}_k(t)=\ov{E}_k(t)/E$, and we refer to the plots of $\en_{k}$ vs.~$k$ and of 
$\overline{\en}_{k}$ vs.~$k$ as the instantaneous and averaged 
modal energy spectrum (m.e.s.), respectively. 
The initial values of the normalized modal energies are then 
$\en_1(0)=1$ and $\en_1(0)\simeq1-\alpha^2E/N$,
for the FPU-$\alpha$ and Toda model respectively, whereas $\en_k(0)=0$ for $k=2,\dots,N-1$. The expected equipartition value of the $\ov{\en}_k$'s is approximately $1/N$. 

\begin{figure}[!ht]
\centerline{
\epsfig{figure=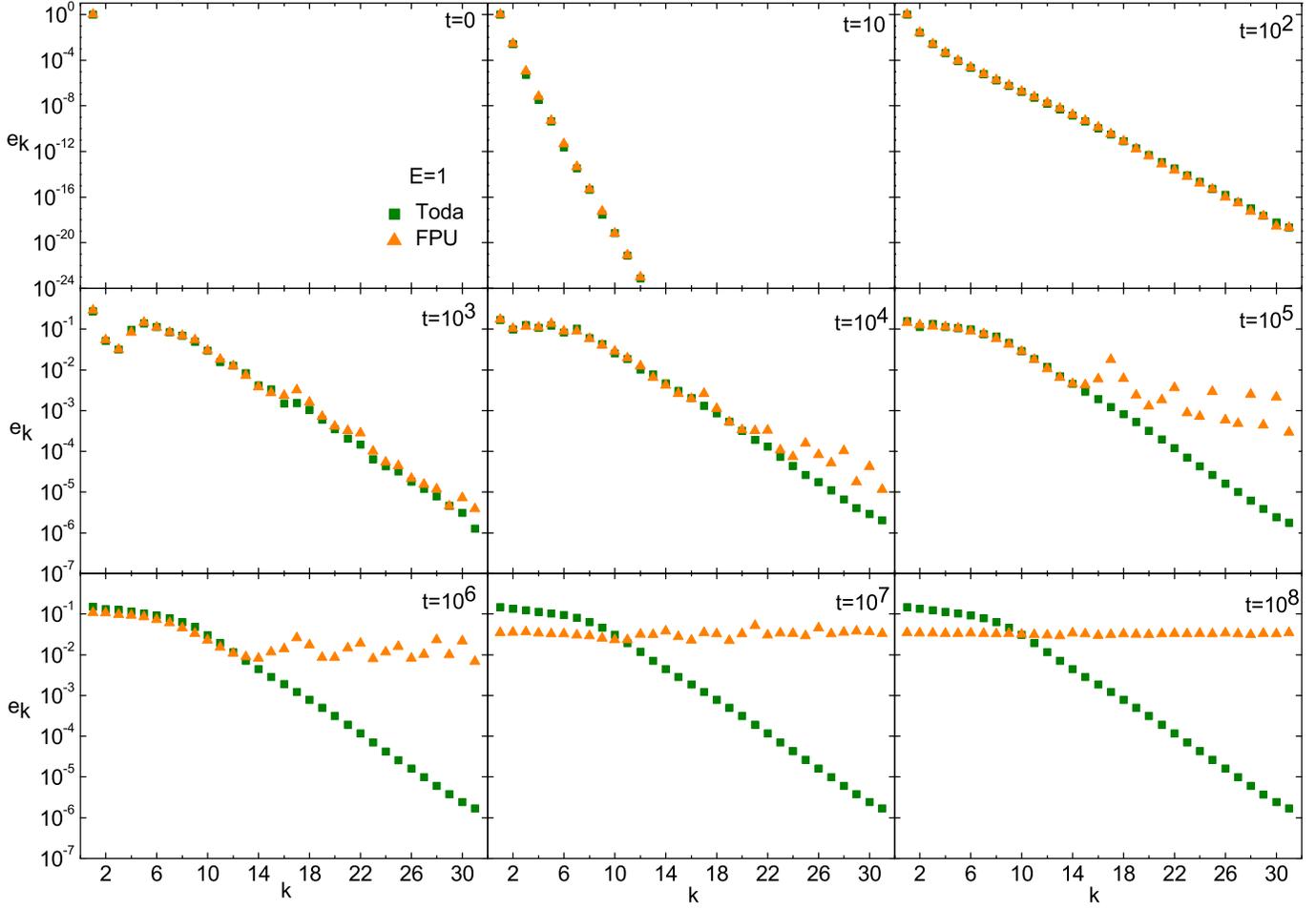,scale=0.65}}
\caption{$\ov{\en}_k$ vs. $k$ (log-linear scale) 
for FPU-$\alpha$ \calleq{H} [orange triangles] and Toda \calleq{HT} [green squares] models, $E=1$, $N=32$, $\alpha=0.33$. From the upper left corner to the lower right one $t=0,10,\dots,10^8$.  Note the change of vertical scale from $t=10^3$ on (last six panels).}
\label{singlerun}
\end{figure}

We report in Fig.~\ref{singlerun} nine snapshots of the FPU and Toda averaged m.e.s. at times $t=0,10,\dots,10^{8}$, all referring to the same run at $E=1$.  
We note  that up to time $t=10^3$ the two spectra almost perfectly
superpose, whereas beginning with $t=10^4$ the tail of the FPU m.e.s. starts to rise while the whole Toda m.e.s. no longer evolves. Finally, at $t=10^8$ the FPU m.e.s. displays an almost perfect flat shape, a signature of statistical mechanical equilibrium. 

\subsection{Stage I: secular avalanche}
On a very short time-scale, the initially excited first mode ($k=1$) looses its energy, while the energies of all the other modes increase with a mode-dependent power of time. 
For both the FPU and the Toda models
the time-evolution of the instantaneous m.e.s.
is described by the following law:
\bea
\en_1(t)&=&1-\frac{\omega_1\omega_2(\mu t)^2}{8}\ ,\lab{e1t}\\
\en_k(t)&=&(\mu t)^{2(k-1)}\ c_k^2\ \ \ \ (2\leq k\leq N-1)\ ,
\lab{ekt}
\eea
where
\be{mu}
\mu\equiv\alpha\sqrt{\frac{E}{N}}\ ,
\ee
is a small parameter  and the time-independent coefficients $c_k$ are recursively computed by solving the nonlinear equations
\bea
c_1=1 \ \ ;\ \ c_{k}=\frac{\omega_k}{4(k-1)}
\sum_{q=1}^{k-1}c_{k-q}c_q \lab{ck}\ \ \ 
(2\leq k\leq N-1)\ .
\eea
These results follow from the smallness of the difference
$\omega_k-k\omega_1$, and essentially consist of a
resonant transfer of energy, where the secular (proportional to $t^2$) growth of the modal energy $\en_2$ in turn implies the law $\en_k\propto t^{2(k-1)}$ ($k\geq2$). 
We call this  stage of the cascade   \emph{the secular avalanche stage} (see Section \ref{secava}). The relevance of the acoustic resonance as a driving mechanism of the cascade in the FPU problem was first 
stressed in Ref. \cite{Ford61}. 

\subsection{Toda equilibrium and quasi-state}

Until a characteristic, energy-dependent time-scale
\be{Tqs}
T^{qs}\approx \frac{N}{\mu}
\ee
the decrease of $\en_1$ and the resonant growth of 
$\en_{k\geq2}$ due to the secular avalanche effectively slow down in the FPU case and actually stop
in the Toda case. The time-averaged m.e.s. of both the FPU and Toda models converge to an exponentially
localized profile, which, for energy values up to about $10^{-2}$, is well described by the expression
\be{qb1}
\ov{\en}_k\simeq k^2\left(\frac{\alpha^2EN^3\ov{\en}_1}{\pi^4}\right)^{k-1}\ov{\en}_1\ ,
\ee
with $\ov{\en}_1$ determined by the 
normalization condition
\be{normh}
\sum_{k=1}^{N-1}\ov{\en}_k=1\ .
\ee
The m.e.s. profile \calleq{qb1}
is that of the $q$-breather associated to mode $k=1$, i.e. the Lyapunov continuation
of the first mode to the nonlinear system \cite{FIK1}. The single energy-dependent frequency
of this periodic orbit, or one-dimensional torus, is 
\be{Om1}
\Omega_{1}(\mu)\simeq\omega_{1}+\mu^{2}\ov{\en}_{1}
\frac{\omega_{1}\omega_{2}}{8(2\omega_{1}-\omega_{2})} + \dots \ 
\ee
Though the agreement between formula \calleq{qb1} and the numerical m.e.s. becomes the less accurate the higher is the energy, we numerically observe that the m.e.s. of both models keep on saturating to (almost) the same exponentially localized profile for all the explored energy values up to $E=1$.
Such an agreement suggests a simple interpretation to the nature of the quasi-state guessed by FPU: 
\emph{the quasi-state of the $\alpha$-model coincides with the final equilibrium state of the Toda model}. It must be stressed that strong indications to such a conclusion
can be found already in Refs. \cite{FFML,Zab06}.

The analytical predictions \calleq{e1t}-\calleq{ekt} and \calleq{qb1}, were obtained by computing a particular resonant normal form Hamiltonian of the system
to leading order.
The same normal form is the starting point of heuristic considerations concerning the observed stability of
the $q$-breather m.e.s. \calleq{qb1} on long times.

\subsection{Stage II: diffusion of tail modes}

After saturation, the FPU m.e.s. slowly continues to evolve and detaches from the reference, stationary Toda m.e.s..
This behavior becomes actually unobservable below some energy threshold $E_c$ ($E_c\simeq0.1$ for $N=32$), at least up to the available observation times ($10^8$ to $10^9$). Above this threshold, there is an evident 
raising of the m.e.s. tail.  
We consider the evolution
of the normalized total energy residing in the last third of the modes, namely
\be{eta}
\eta(t)\equiv\sum_{k=22}^{31}\en_k(t)\ ,
\ee
The choice of the last third of modes in defining the tail of the m.e.s. is of empirical nature, and agrees, when the quasi-state is well described by the $q$-breather, with the definition of tail given in Ref. \cite{PF} and based on resonance arguments. We find that, after a transient time approximately coinciding 
with the time-scale $T^{qs}$, the quantity $\ov{\eta}(t)$
starts to increase with time as a power law
\be{diff}
\ov{\eta}(t)\sim Dt^{\gamma}\ ,
\ee
within an energy depending time-window. 
Both the exponent $\gamma$ and the coefficient $D$ depend
on the energy. The diffusion exponent $\gamma(E)$ roughly displays a step-like behavior: $\gamma(E)\simeq0$ for 
$E\ll E_{c}$ and $\gamma(E)\simeq1$ as $E\gg E_{c}$.
The energy dependence between the two limits is not smooth: the exponent $\gamma(E)$ strongly fluctuates around $E_c$, the latter value being more properly identified as the center of a transition interval rather than an actual sharp threshold. This is possibly a signature of anomalous diffusion processes, characterizing the dynamics of tail modes (recall that 
$\en_k=E_k/E$ depends quadratically on the modal coordinates $P_k$ and $Q_k$). Depending on the value of the energy $E$
one can observe either sub-diffusion ($\gamma<1$) or super-diffusion ($\gamma>1$). An analytical derivation of the law \calleq{diff} is currently missing.

We then extrapolated the time-scale to equipartition $T^{eq}$, defined as the time necessary for 
its time average $\ov{\eta}$ to reach its approximate equipartition value $1/3$:
\be{Teq}
T^{eq}\equiv(3D)^{-1/\gamma}\ .
\ee
Both $D$ and 
$\gamma$ can be therefore obtained on relatively short times. These  values lead to lower bounds for the actual equipartition time, which for example can reach the unconceivable value of 
$10^{1000}$ at $E=0.01$.  We find that $T^{eq}\propto E^{-a}$
at higher energies (from $E=0.5$ up to $2$), with $a$ close to $3$, whereas $T^{eq}\sim\exp(c/E^{b})$ for low energies
(in the range $0.01$ to $0.1$), with $b\simeq0.7$. 

In this second stage of the dynamics we also measured the \emph{erosion} of one of the first integrals of the Toda model independent of the Hamiltonian, computing it along the FPU dynamics.  It turns out that, above the energy threshold $E_c$, 
when the diffusive-like raising of the tail modes becomes observable, such a quantity undergoes a drift starting at times of the order of $T^{qs}$. This behavior indicates that the relaxation to equilibrium of the FPU system is characterized by the drift of the Toda integrals, which become quasi-invariants of the FPU system at low energy.

\section{Phenomenology and Numerics}

We now come to a detailed description of the phenomenology of the problem, presenting our numerical results and their comparison with the theoretical predictions.

\subsection{Stage I: secular avalanche}
\label{secava}

\begin{figure}[!ht]
\centerline{\epsfig{figure=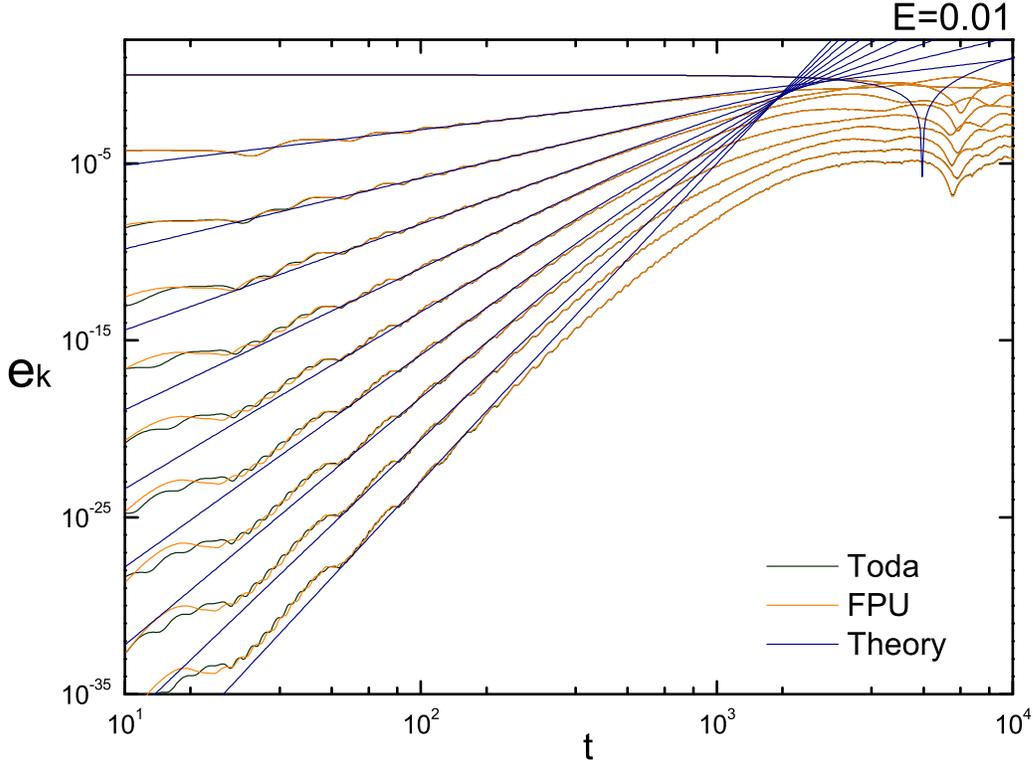,scale=0.6}}
\caption{$\en_{k}(t)$ vs. $t$ (log-log scale) for $k=1,\ldots,10$ at $E=0.01$; black
and orange curves refer respectively to the Toda and the FPU numerically computed energies, while blue lines correspond to the theoretical law 
\calleq{e1t}-\calleq{ekt}.}
\label{ek_t001}
\end{figure}

The secular avalanche stage is shown in Fig.~\ref{ek_t001}, where
the log-log scale plot of $\en_k(t)$ vs. $t$ at $E=0.01$ 
is reported, both for the Toda and the FPU-$\alpha$ system, together with the lines corresponding to the
theoretical law \calleq{e1t}-\calleq{ekt}.
The values of the mode number are restricted to $k=1,\dots,10$
just to improve the readability of some details in the figure.
One can note the almost perfect superposition, mode by mode, of the Toda and FPU numerical modal energies, which are both well described, on the very short term, by the law 
\calleq{e1t}-\calleq{ekt}. The same good agreement of the latter theoretical law with the numerical data is found all over the energy range $10^{-2}$ to $1$, though further figures are not reported.

The law $\en_{k}(t)\propto t^{2(k-1)}$ for low
modes is due to the smallness of the difference
$k\omega_{1}-\omega_{k}$, i.e. to their almost complete resonance. Indeed, mode $k=2$ is forced in resonance by
the second harmonic of mode $k=1$, so that its amplitude grows
linearly and its energy grows quadratically with time: 
$\en_{2}\propto t^{2}$. Mode $k=3$ is in turn forced in resonance by the product of amplitudes of modes $k=1$ and $k=2$, which
has a factor $t$ in front, so that $\en_{k}\propto(t^{2})^{2}=t^{4}$, and so on. In this way a resonant cascade of energy transfer sets in,
which causes the growth of the initially unexcited modes one after the other. Notice that for high modes, the difference 
$\omega_{k}-k\omega_{1}$ is not small, but their amplitude is
initially zero, and this (as shown in Section IV) ensures the validity, also for them, of the approximation leading to the secular avalanche law \calleq{e1t}-\calleq{ekt}.

Let us denote by $T^{sa}_{k}(E,N)$ the mode-dependent characteristic time of validity of the secular avalanche law \calleq{e1t}-\calleq{ekt}. For the first mode this time can be estimated by simply requiring that the leading order estimate \calleq{e1t} yields $\en_{1}\geq0$. Setting 
$\en_1=0$ in \calleq{e1t} one gets $T^{sa}_{1}=2\sqrt{2}/(\mu\sqrt{\omega_{1}\omega_{2}})\simeq2N/(\pi\mu)$. 
Moreover, in the regime of weak mode-coupling explored here, almost all the energy is retained by the harmonic part of the Hamiltonian, so that the sum $\sum_{k=1}^{N-1}\en_{k}$
is close to its initial value (one) for all times, which in turn implies $\en_{k}\leq1$ for all $k$'s. Setting $\en_k=1$ in
\calleq{ekt} one gets 
$T^{sa}_k=c_k^{-\frac{1}{k-1}}/\mu$ for $k\geq2$. We finally notice that $T^{sa}_{k}\sim T^{sa}\equiv 
0.2\ N/\mu$ for large values of $k$. By exploiting such asymptotic behavior, the law \calleq{ekt} reads, for $k\gg1$, $\en_k(t)=
e^{-2(k-1)\sigma(t)}$, where $\sigma(t)=-\ln(t/T^{sa})$. 
A logarithmic time-dependence of the slope $-\sigma(t)$ of the m.e.s on the short term was first predicted, with methods based on continuum approximations, in Refs. \cite{Fetal,LPRV}.
It is interesting to notice that the time-scale 
\calleq{Tqs}, derived below (in Section \ref{roleres}) from the point of view of mode dynamics, was first determined as the break-down time of the solution of the nonlinear wave-equation approximating the FPU dynamics in the continuum, zero dispersion limit.\cite{Zab62,KZ64}

\subsection{Toda equilibrium and quasi-state}

\begin{figure}[!ht]
\centerline{
\epsfig{figure=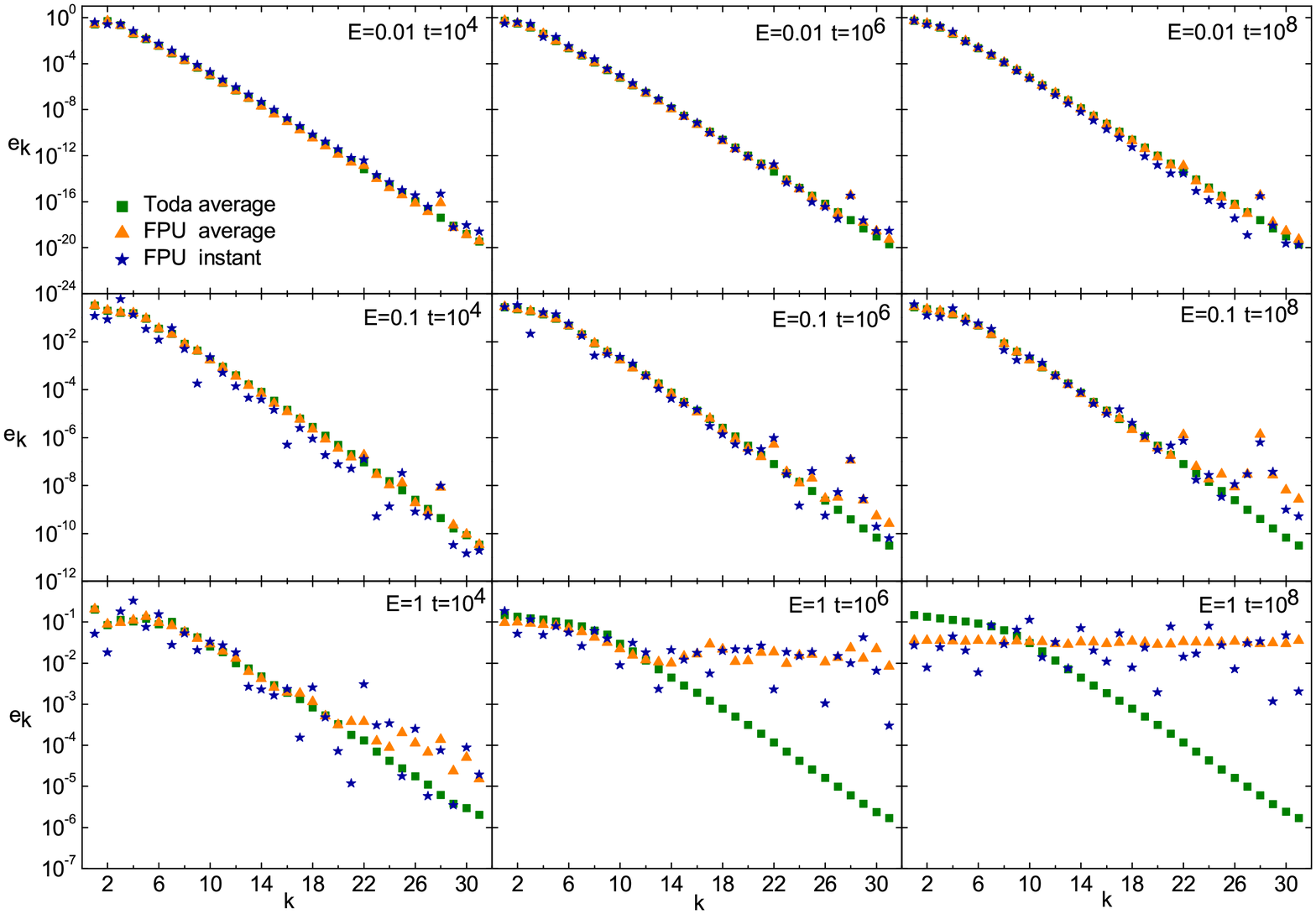,scale=0.65}}
\caption{$\ov{\en}_k(t)$ vs. $k$ (log-linear scale) for the Toda (green squares) and the FPU-$\alpha$ (orange triangles) systems, and $\en_k(t)$ vs. $k$ for the FPU-$\alpha$ model (blue stars). $E=0.01,0.1,1$ from top to bottom, $t=10^4,10^6,10^8$ from left to right, i.e.~the energy is constant along the rows and the time is constant along the columns. Notice the change of vertical scale from top to bottom.}
\label{multirun}
\end{figure}

At the end of the secular avalanche, the time-averaged m.e.s.
($\ov{\en}_k$ vs. $k$) of both the FPU-$\alpha$ and the Toda models saturate to one and the same exponentially localized profile - the FPU quasi-state. For larger times the Toda time-averaged m.e.s. no longer evolves, whereas the FPU one keeps on evolving, the more slowly the lower is the value of the energy 
$E$. In order to illustrate this, in Fig.~\ref{multirun}
we report both the time-averaged Toda and FPU m.e.s.
and the FPU instantaneous one (log-linear scale). 
The nine panels refer to three values of the energy $E=0.01,0.1,1$
at times $t=10^4,10^6,10^8$. The value of the energy $E$ is fixed in each row,
whereas that of the time is fixed in each column. At $E=0.01$ (first row) the Toda and FPU spectra are almost perfectly superposed at any time, the only observable difference being the small fluctuations
$\ov{\en}_k-\en_k$ of the FPU energies of the last (highest) few modes. 
At $E=0.1$ (second row) the FPU spectral fluctuations
$\ov{\en}_k-\en_k$ become observable for any $k$, and increase with time. Also
at $t=10^6$ the tail of the FPU m.e.s. starts to rise
detaching from that of the Toda. Finally, at $E=1$ (third row) (see also Fig.~\ref{singlerun}) one observes, in the FPU m.e.s.,
even larger fluctuations, the raising of the tail and the relaxation to equipartition. 

Fig.~\ref{multirun} also shows that at low energy, when the FPU model takes a long time to relax to equilibrium, its motion stays close to the invariant torus of the Toda model corresponding to (almost) the same initial condition. 
The slow variation along the FPU actual trajectory of the Toda invariants is at the basis of the FPU paradox.
This was first conjectured in Ref. \cite{FFML}, where a phenomenological picture of the Toda dynamics based on inverse scattering is presented. 
In particular, in the latter reference, it is shown that at small energy, the Toda dynamics corresponding to the initial excitation of a single mode takes place on a one-dimensional torus, and that
such a periodic orbit must be the Lyapunov continuation to the nonlinear problem of the linear normal mode solution.

\begin{figure}[!ht]
\centerline{
\epsfig{figure=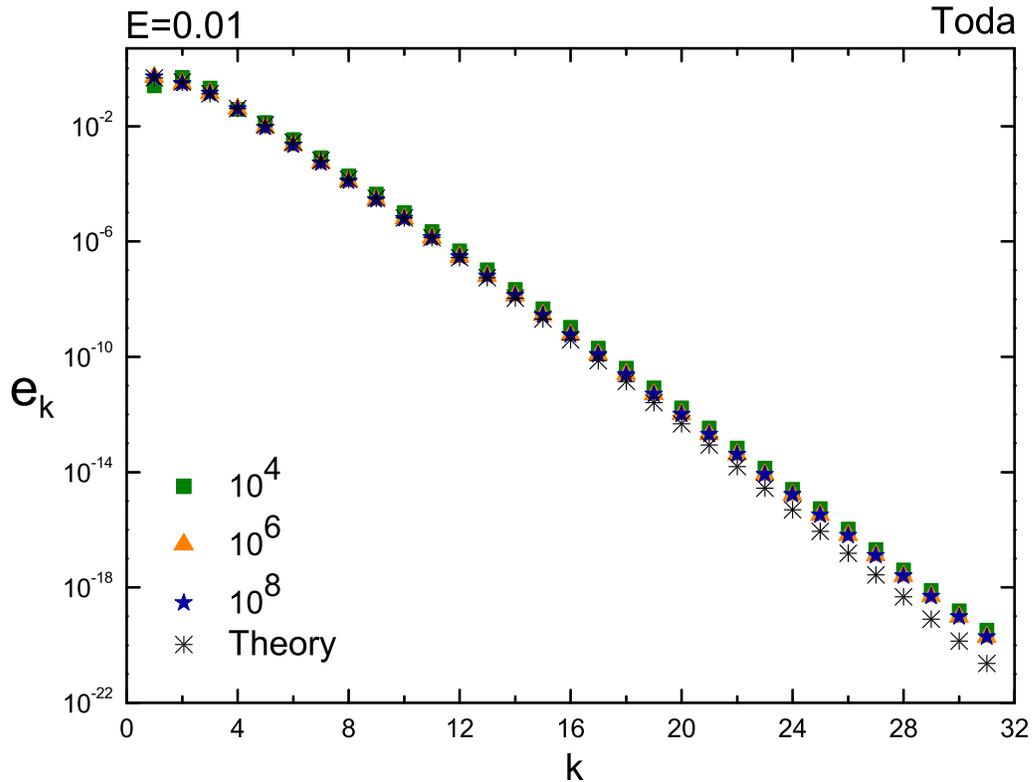,scale=0.6}}
\caption{Time-averaged Toda m.e.s. at three different times ($10^4,10^6,10^8$), $E=0.01$. The asterisks correspond to the theoretical prediction \calleq{qb1}.}
\label{ek_kT}
\end{figure}

With this in mind, in Fig.~\ref{ek_kT} the time-averaged m.e.s. (in log-linear scale) of the Toda model at $E=0.01$ is reported for three different values of the time $t=10^4,10^6,10^8$, using different symbols.
Notice that the three numerical spectra at different times almost perfectly superpose, which shows that the time-averaged m.e.s. we are looking at is stationary.
The theoretical m.e.s. \calleq{qb1} (with $\ov{\en}_{1}$ computed from \calleq{normh}) is also plotted in Fig.~\ref{ek_kT}
(asterisks). The agreement with the numerical data is very good
and holds for energies lower than $E=0.01$. On the other hand, in Ref. \cite{FFML} it is also shown that the dimension of the Toda torus corresponding to the a single mode excitation grows with the energy, so that the (Toda) motion becomes quasi periodic. This explains why the $q$-breather m.e.s. \calleq{qb1} fails to describe well, at larger energy values, the numerical spectra of the Toda and FPU systems, although the latter spectra go on to
almost superpose on short times. The analytic description of the Toda or FPU m.e.s. corresponding to the full quasi-periodic case is beyond the scope of the present paper and is left to future investigation (see however Refs. \cite{CEB1,CEB2} for interesting results in this direction).

\subsection{Stage II: diffusion of tail modes}

\begin{figure}[!ht]
\centerline{
\epsfig{figure=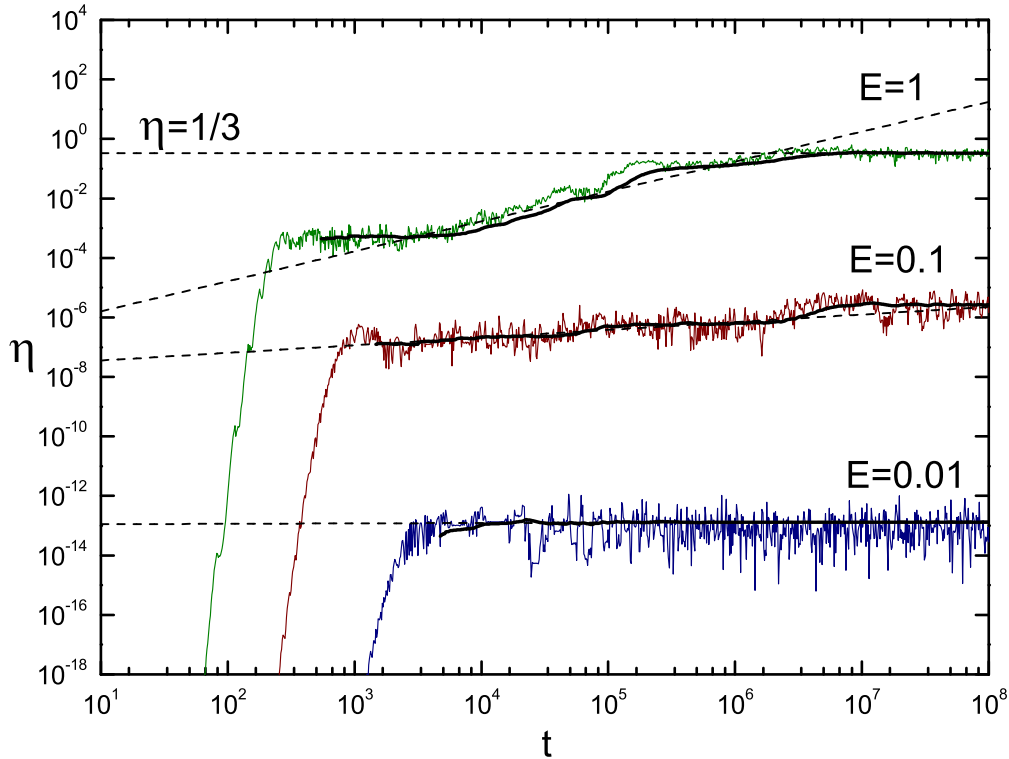,scale=0.6}
}
\caption{Bottom to top: $\eta(t)$ vs. $t$ (log-log scale) for $E=0.01$ (blue) $0.1$ (magenta) and $1$ (green), together with the corresponding time-averages $\ov{\eta}(t)$ vs. $t$ (black curves) and the fitting lines $Dt^\gamma$ vs. $t$ (dashed lines).}
\label{eta_t}
\end{figure}

\begin{figure}[!ht]
\centerline{
\epsfig{figure=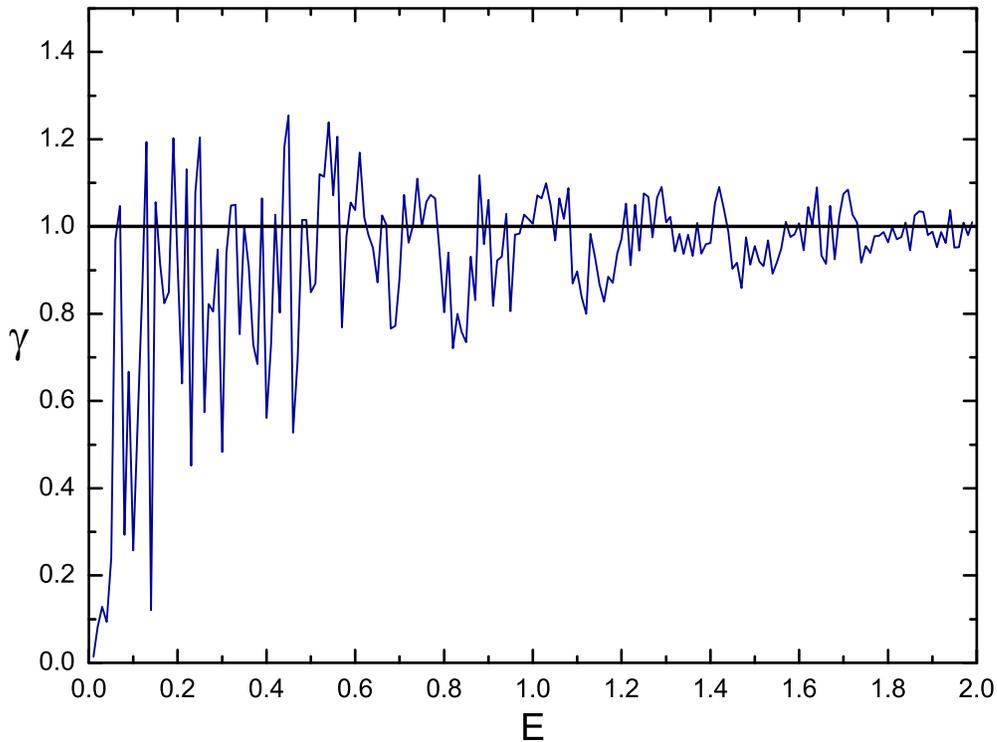,scale=0.6}
}
\caption{$\gamma$ vs. $E$ corresponding to $\ov{\eta}$.}
\label{gamma_E}
\end{figure}

\begin{figure}[!ht]
\centerline{
\epsfig{figure=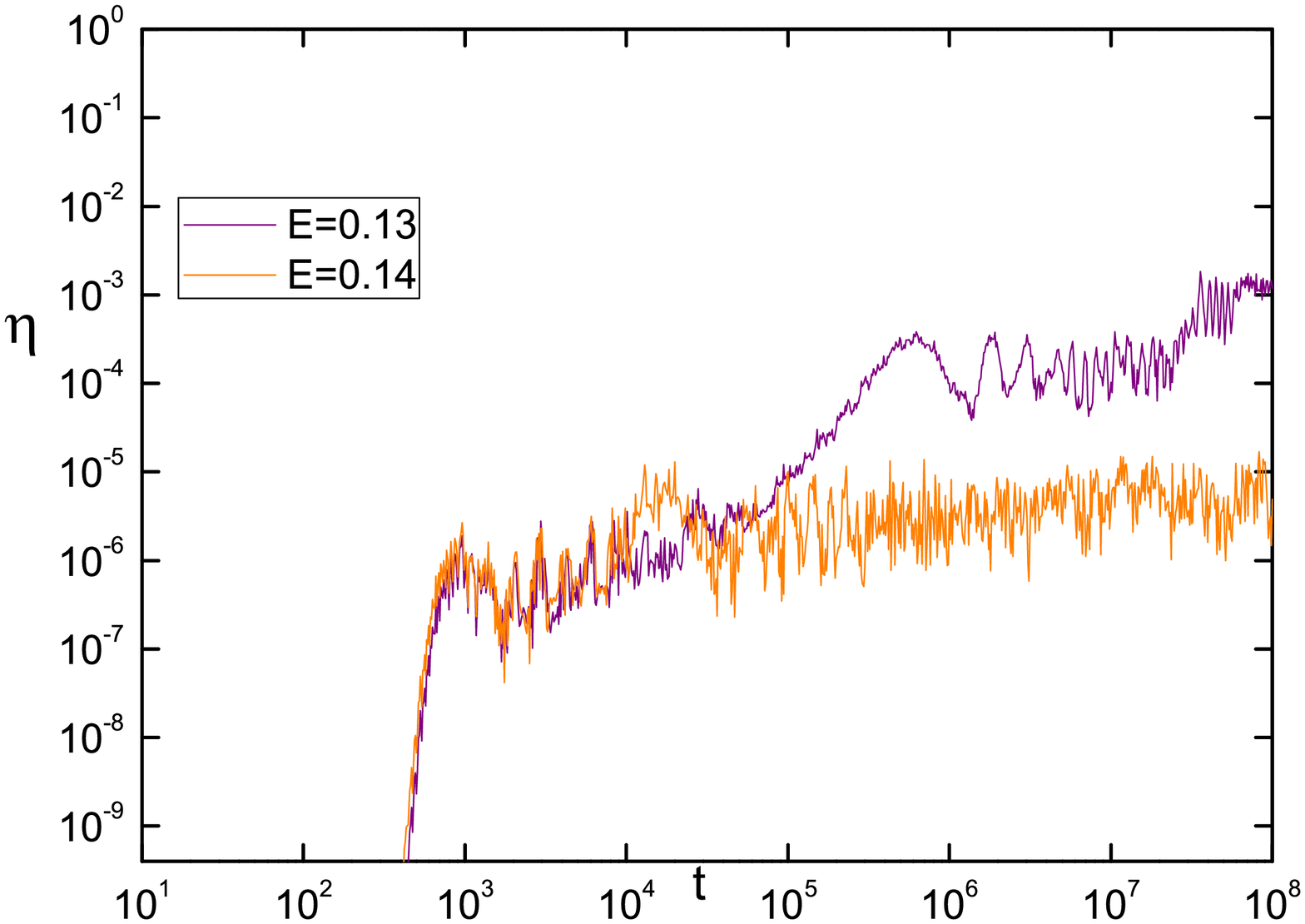,scale=0.6}
}
\caption{$\eta(t)$ vs. $t$ for two values of the energy differing
by $\Delta E=10^{-2}$: the upper and lower curves correspond to the energy values $E=0.13$ and $E=0.14$, respectively.}
\label{eta_tcomp}
\end{figure}

In order to describe the approach to equipartition in the FPU system, in Fig.~\ref{eta_t} we plot (in log-log scale) the instantaneous normalized tail energy
$\eta(t)=\sum_{k=22}^{31}\en_k(t)$ and the corresponding 
time-averaged values $\ov{\eta}(t)$ vs. $t$,
for three values of the energy $E=0.01,0.1,1$
(bottom to top). Note that the instantaneous values evolve by fluctuating around the corresponding time-averages. 
We observe a growth of both quantities, well fitted by a 
power-law $\ov{\eta}(t)\sim Dt^\gamma$
within some time-interval $[t_{0},t_{1}]$. It follows the linear
behavior $\log\ov{\eta}(t)\simeq\log D+
\gamma\log t$ in log-log scale, as reported in Fig.~\ref{eta_t}.
Notice that when $E=1$, $\eta$ and $\ov{\eta}$ asymptotically reach the equipartition value $1/3$ and do not grow further, as expected.

By a systematic and detailed numerical investigation we first determined the time intervals $[t_{0}(E),t_{1}(E)]$
and then computed $\gamma(E)$ and $D(E)$ by a least squares 
fit (200 values of $E$ were considered, in the energy interval $0.01$ to $2$, with an energy step $\Delta E=0.01$).
The errors on both quantities turn out to be small (a few percents of the value) except for very low energy values, close to $10^{-2}$, where large errors 
affect the almost vanishing exponent $\gamma$. The function $\gamma(E)$ is plotted in 
Fig.~\ref{gamma_E}. Despite strong fluctuations, we find that
$\gamma$ changes from a value very close to zero at energies lower than $0.1$ to values oscillating around $\gamma=1$ for energies larger than $0.1$. 
Moreover, the absolute value of the fluctuations decreases with increasing energy.

We conjecture that the fluctuations in the exponent $\gamma$ are intrinsic to the dynamics, and 
are due to the complicated phase space structure of the FPU system. To show that, we plot $\eta(t)$ in Fig.~\ref{eta_tcomp} for two values of the energy which differ by
$\Delta E=10^{-2}$. Indeed, the growth rates differ substantially:
the values of the exponent $\gamma$ corresponding to these two energies are $\gamma(0.14)=0.03$ and $\gamma(0.13)=1.1$.

\subsubsection{Estimate of the time to equipartition}

\begin{figure}[!ht]
\centerline{
\epsfig{figure=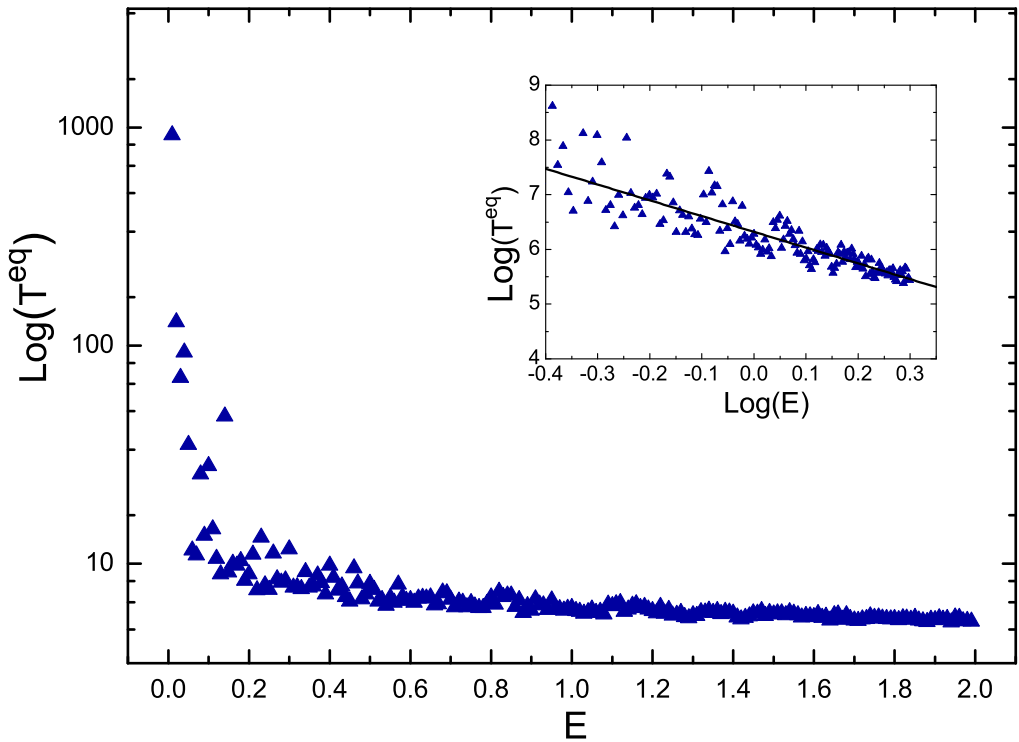,scale=0.6}
}
\caption{Estimated $\log(T^{eq})$ vs. $E$ from average $\ov{\eta}$ data. Inset:
$\log(T^{eq})$ vs. $\log(E)$. The line drawn has slope $-2.9$.}
\label{Teq_E32}
\end{figure}

The observed power-law growth $\ov{\eta}\sim Dt^\gamma$
allows us to define the extrapolated equipartition time such that
$Dt^\gamma$ equals $1/3$, i.e. the equipartition value of 
$\ov{\eta}$, which yields $T^{eq}\equiv(3D)^{-1/\gamma}$. In 
Fig.~\ref{Teq_E32} we plot 
$\log(T^{eq})$ vs. $E$. We observe a sharp crossover around $E=0.1$, below which $T^{eq}$
increases dramatically, up to values 
$T^{eq}\approx 10^{1000}$. Such large times strongly suggest that below the reference energy threshold 
$E_c\simeq0.1$,
the quasi-state of the FPU system (i.e. the equilibrium state of the Toda model) might become stable in the sense of Nekhoroshev, i.e. over exponentially long times of the order
$10^{c/E^{b}}$, with suitable constants 
$c$ and $b$. We get a rough estimate for the stretching exponent 
$b\simeq 0.7\pm0.2$. Existing numerical results in the 
literature\cite{CCPC,DLR,BGP,BP11} suggest one that for larger values of the energy the time to equipartition grows inverse proportional to a power of the energy $E$.
In the inset of Fig.~\ref{Teq_E32} we plot
$\log(T^{eq})$ vs. $\log(E)$ in the energy interval $0.5$ to $2$. 
We indeed find some evidence for a power law, with exponent 
$a\simeq2.9$. This number is close to the value 
$a=3$ first reported in Ref. \cite{CCPC} for $N=32$, and there obtained with other methods.

\subsubsection{Evolution of a Toda integral}

\begin{figure}[!ht]
\centerline{
\epsfig{figure=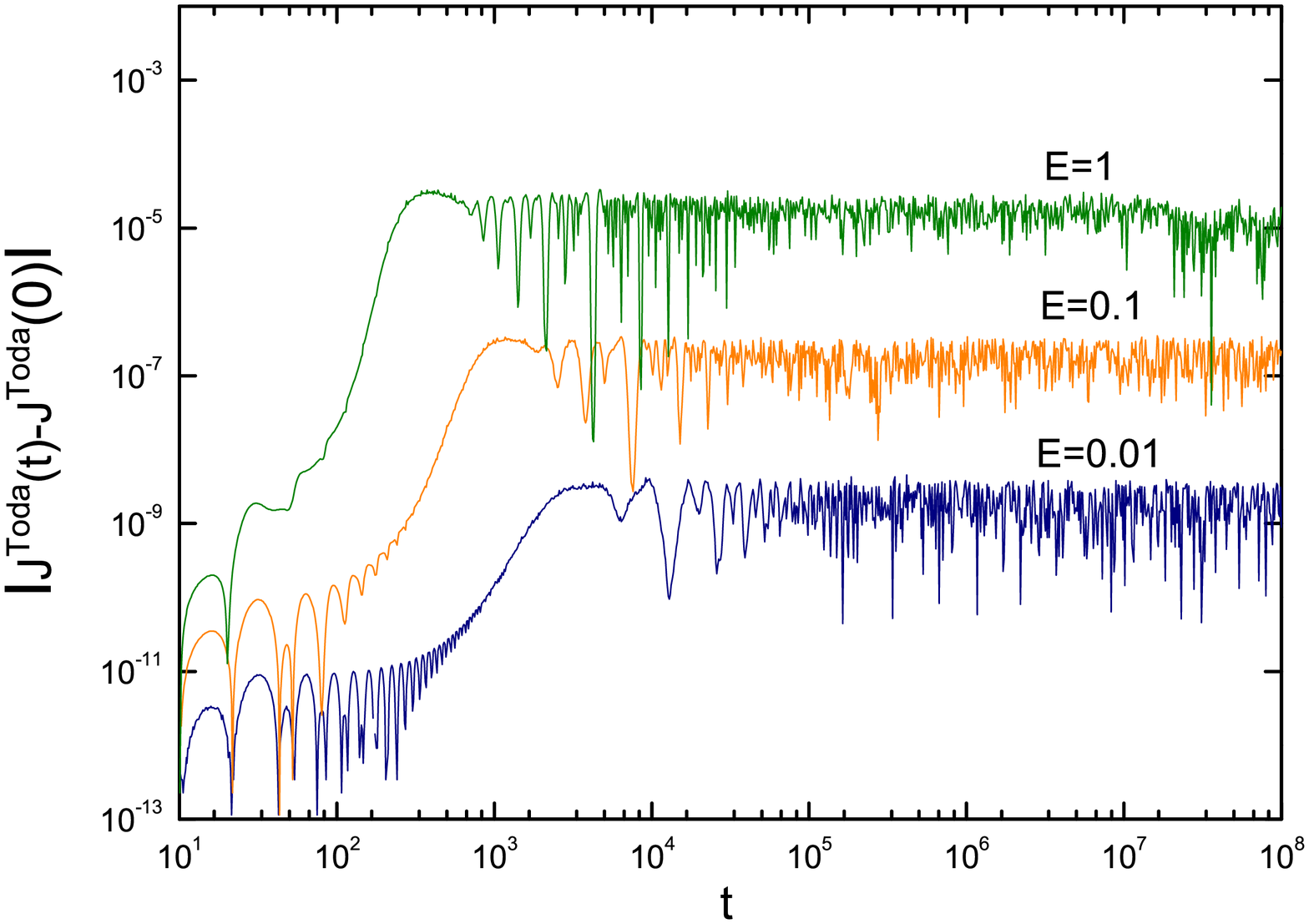,scale=0.6}
}
\caption{$|J^{Toda}(t)-J^{Toda}(0)|$ vs. $t$ (log-log scale) at $E=0.01$ (blue),
$E=0.1$ (orange) and $E=1$ (green).}
\label{JerrT}
\end{figure}

\begin{figure}[!ht]
\centerline{
\epsfig{figure=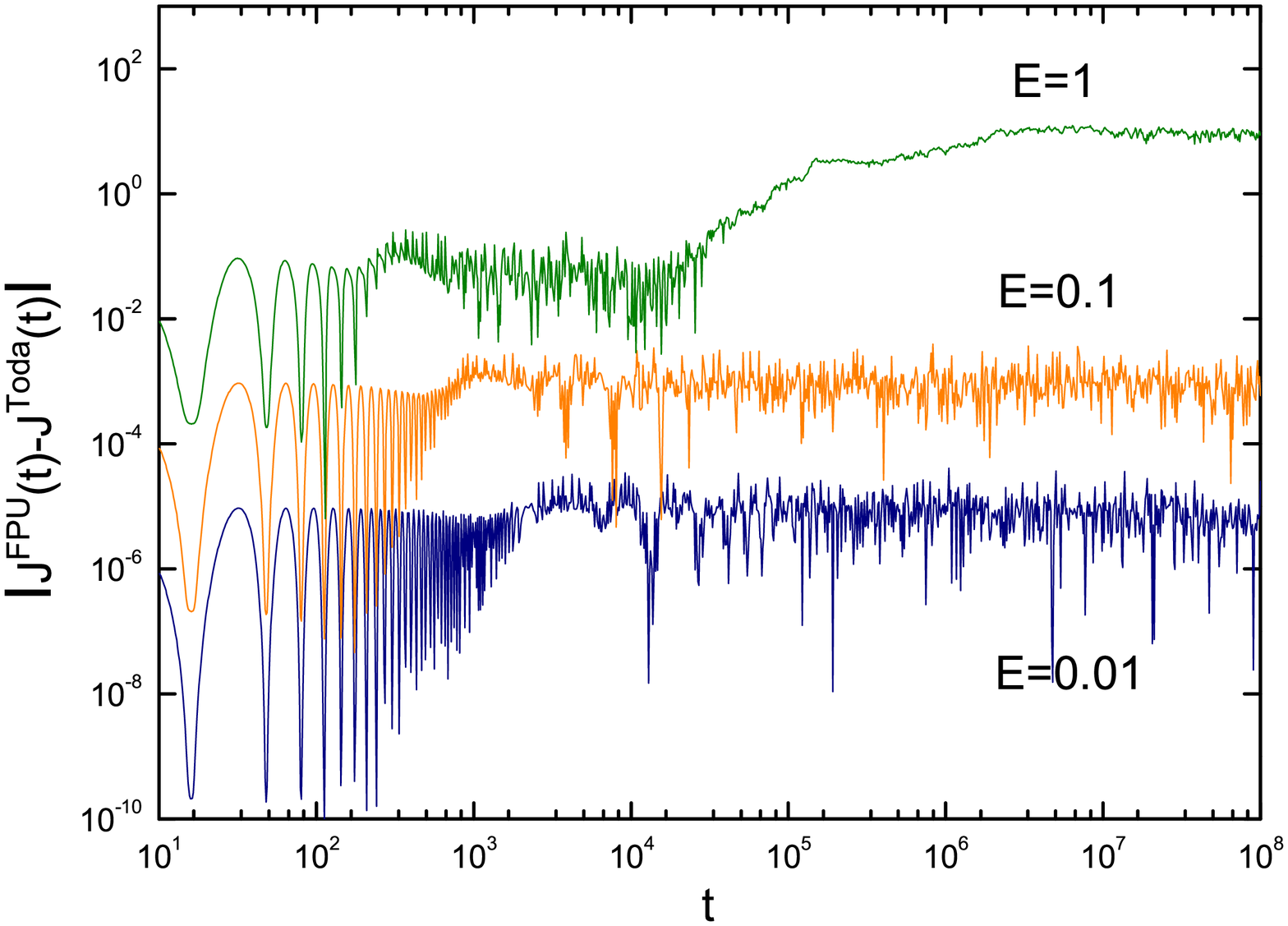,scale=0.6}
}
\caption{$|J^{FPU}(t)-J^{Toda}(t)|$ vs. $t$ (log-log scale) at $E=0.01$ (blue),
$E=0.1$ (orange) and $E=1$ (green).}
\label{Jdiff}
\end{figure}

We have studied the evolution of one of the additional constants of motion of the Toda model. The integral $J(q,p)$ we considered is \cite{He}
\be{J}
J\equiv\sum_{n=0}^{N-1}\left[
\frac{p_n^4}{2}+\frac{(q_n^2+q_{n+1}q_n+q_{n+1}^2)f_n}{\alpha}
+\frac{(f_{n+1}+f_n+f_{n-1})f_n}{4\alpha^{2}}
\right]\ ,
\ee
where $f_n\equiv e^{2\alpha(q_{n+1}-q_n)}/(2\alpha)$
($n=0,\dots,N-1$) are the Flaschka variables \cite{Fl}, with boundary conditions $f_{-1}=f_0$, $f_N=f_{N-1}$, inherited by the restriction of the $2N$-periodic case to that of fixed ends. 
Since $J$ is an integral of the Toda model,
$\dot{J}=\{J,H_T\}=0$ along the Toda flow. Along the FPU flow the quantity 
$\dot{J}=\{J,H_{\alpha}\}=
\{J,H_{\alpha}-H_{T}\}$ is expected to be small in the perturbative regime where the two models can be considered to be close to each other. Let us denote by $J^{FPU}(t)$
and by $J^{Toda}(t)$ the evolution of the quantity \calleq{J} along the \emph{numerical} FPU and Toda trajectories, respectively.

In Fig.~\ref{JerrT} the absolute error $|J^{Toda}(t)-J^{Toda}(0)|$ vs. $t$ is reported, for the three different values of the energy $E=0.01,0.1,1$ (bottom to top), in log-log scale. 
After a transient time, the error stabilizes quickly at a 
practically constant value ranging from $10^{-9}$ at $E=0.01$ to $10^{-5}$ at $E=1$. Thus the numerical algorithm reproduces the integrable Toda dynamics with very good accuracy. 

In Fig.~\ref{Jdiff} we  plot the evolution of 
$|J^{FPU}(t)-J^{Toda}(t)|$, for the same three values of the energy $E=0.01,0.1,1$ (bottom to top), in log-log scale.
For $E=0.01$ or $0.1$ this  difference stays almost constant (up to the final observation time $10^8$), 
while for $E=1$ it starts to grow 
at about $t=10^4$ to eventually saturate at a constant value.
Therefore at low energies the FPU system
can be regarded as a perturbation of the Toda one, the exact integrals of the latter being adiabatic invariants, or quasi-integrals, of the former. Note 
that at $E=0.01$ and $0.1$ no effective tendency
to equipartition was observed up to $t=10^8$. On the other hand,
for $E=1$ the tail of the FPU m.e.s. starts to effectively raise at $t=10^4$, when the drift of the Toda integral starts up.

\section{Theory}

Our theoretical approach is based on Hamiltonian perturbation theory. 
We construct a resonant normal form Hamiltonian of the FPU and Toda systems. The laws \calleq{e1t}-\calleq{ekt}
are then derived by solving the normal form equations, to leading order, for short times. 
The law \calleq{qb1} is instead obtained as the m.e.s. of a time-periodic solution of the normal form equations. The theoretical explanation of the diffusion law \calleq{diff} goes beyond our present capabilities.

\subsection{Complex modal variables}

Let us consider the Hamiltonian \calleq{H} of the FPU 
$\alpha$-model. We first perform a change of variables
passing from the real space coordinates 
$(q,p)\in\mathbb{R}^{2(N-1)}$ to the complex modal variables 
$u\in\mathbb{C}^{N-1}$. The complex modal variables $u_{k}$ are defined in terms of the real ones by
\be{uk}
u_k\equiv\frac{\omega_kQ_k+iP_k}{\sqrt{2E}}\ \ ,\ \ 
|u_k(t)|^2=\frac{E_k(t)}{E}=\en_k(t)\ ,
\ee
whose utility is self-explanatory. The FPU Hamiltonian
\calleq{H}, when expressed in the
new complex variables $u_{k}$, reads
\be{Hmodes}
\mathcal{H}(u,u^*)=
\underbrace{\sum_{k=1}^{N-1}|u_k|^2}_{\mathcal{H}_{2}}
+\underbrace{\frac{\mu}{12}
\sum_{k_1,k_2,k_3=1}^{N-1}\Delta_{k_1,k_2,k_3}
\prod_{j=1}^3(u_{k_j}+u_{k_j}^*)}_{\mathcal{H}_{3}}\ ,
\ee
where a superscript asterisk denotes complex conjugation, and the mode-coupling coefficient
\be{Delta}
\Delta_{k_1,k_2,k_3}\equiv
\delta_{k_1+k_2,k_3}+\delta_{k_2+k_3,k_1}+\delta_{k_3+k_1,k_2}-\delta_{k_1+k_2+k_3,2N}
\ee
has been introduced ($\delta_{n,m}$ is the usual Kronecker symbol). We also recall that
\be{mu2}
\mu\equiv\alpha\sqrt{\frac{E}{N}}
\ee
denotes a small parameter in the theory. In \calleq{Hmodes}
we have explicitly pointed out the quadratic part $\mathcal{H}_{2}$
and the cubic part $\mathcal{H}_{3}$ of the Hamiltonian. The Hamilton equations of motion $\dot{u}_k=\{u_{k},\mathcal{H}\}=
-i\omega_k\partial\mathcal{H}/\partial u_k^*$
associated to the Hamiltonian $\mathcal{H}$ explicitly read
\be{eqn}
\dot{u}_k=-i\omega_{k}\left[u_{k}+\frac{\mu}{4}
\sum_{p,q=1}^{N-1}\Delta_{k,p,q}(u_{p}+u_{p}^{*})
(u_{q}+u_{q}^{*})\right]\ .
\ee
Finally, the initial condition \calleq{incond} expressed in the new variables reads 
\be{indat}
u_k(0)=\delta_{k,1}\ .
\ee

\subsection{Averaging}

The initial condition \calleq{indat}, at low energy, is expected to excite an oscillation of the first mode at a 
frequency $\Omega_{1}$ close to the unperturbed one, the difference $\Omega_{1}-\omega_{1}$ 
vanishing as $\mu\rightarrow0$. Indeed, in the latter limit, equations \calleq{eqn} become linear
and are solved by $u_{k}=e^{-i\omega_{1}t}\delta_{k,1}$. 
Now, by substituting the leading order ansatz 
$u_{k}=e^{-i\Omega_{1}t}\delta_{k,1}$ in the right hand side of equations \calleq{eqn}, and making use of \calleq{Delta}, one realizes by a direct inspection that the mode $k=2$ is forced, among others, by a term oscillating at a frequency 
$2\Omega_{1}\simeq2\omega_{1}\simeq\omega_{2}$. 
Thus, mode $k=2$ responds close to resonance and its amplitude
grows, which requires to modify the starting ansatz to
$u_{k}=Ae^{-i\Omega_{1}t}\delta_{k,1}+Be^{-i2\Omega_{1}t}
\delta_{k,2}$. Re-inserting the latter in the right hand side of
\calleq{eqn} one finds that mode $k=3$ is almost resonantly pumped at a frequency $3\Omega_{1}\simeq3\omega_{1}
\simeq\omega_{3}$, and so on. Such a process leads to an almost resonant transfer of energy from the fundamental mode to 
higher frequency modes, with mode $k$ oscillating at the frequency $k\Omega_{1}$. 
For this reason, we perform the time-dependent change of variables $(u,u^{*})\mapsto(v,v^{*})$ 
\be{vtou}
(u,u^{*})=\Phi^{\Omega_{1}t}(v,v^{*})\ :\ u_{k}=
e^{-ik\Omega_{1}t}v_{k}
\ee
for $k=1,\dots,N-1$. The above defined map
$\Phi^{\Omega_{1}t}$ is actually the $2\pi/\Omega_{1}$-periodic flow of the Hamiltonian
\be{hOm1}
h_{\Omega_{1}}(u,u^{*})=\sum_{k=1}^{N-1}
\left(\frac{k\Omega_{1}}{\omega_{k}}\right)|u_{k}|^{2}\ .
\ee
Notice also that all the functions of the 
modal energies $\en_{k}=|u_{k}|^{2}$, such as $\mathcal{H}_{2}$
and $h_{\Omega_{1}}$, are invariant under the  flow
$\Phi^{\Omega_{1}t}$, i.e. Poisson-commute with 
$h_{\Omega_{1}}$. The new variables $v$ evolve according to the Hamilton equations $\dot{v}_{k}= \{v_{k},\mathcal{A}\}=
-i\omega_{k}\partial \mathcal{A}/\partial v_{k}^{*}$
associated to the explicitly time-dependent,
$2\pi/\Omega_{1}$-periodic Hamiltonian $\mathcal{A}$:
\be{A}
\mathcal{A}(v,v^{*},\Omega_{1}t)\equiv
\mathcal{H}_{2}(v,v^{*})-h_{\Omega_{1}}(v,v^{*})+
\mathcal{H}_{3}\left(\Phi^{\Omega_{1}t}(v,v^{*})\right)\ .
\ee
According to the usual averaging theory \cite{BM,SVM}, we replace the explicitly time-dependent Hamiltonian $\mathcal{A}$ with
its time average $\ov{\mathcal{A}}$. This amounts to replacing 
$\mathcal{H}_{3}\circ\Phi^{\Omega_{1}t}$, on the right hand side of \calleq{A}, with its time average
\be{H3avform}
\ov{\mathcal{H}}_{3}(v,v^{*})=\frac{\Omega_{1}}{2\pi}
\int_{0}^{\frac{2\pi}{\Omega_{1}}}
\mathcal{H}_{3}\left(\Phi^{\Omega_{1}s}(v,v^{*})\right)\ ds\ .
\ee
We then obtain an averaged system, whose solution stays close to that
of the original system over times inverse proportional to the small parameter. 
The averaged Hamiltonian $\ov{\mathcal{A}}=\mathcal{H}_{2}-
h_{\Omega_{1}}+\ov{\mathcal{H}}_{3}$ of the system then reads
\be{Aav}
\ov{\mathcal{A}}(v,v^{*})=\sum_{k=1}^{N-1}\left(
\frac{\omega_{k}-k\Omega_{1}}{\omega_{k}}\right)|v_{k}|^{2}+
\frac{\mu}{4}
\sum_{p,q=1}^{N-1}(v_{p+q}^*v_pv_q+
v_{p+q}v_p^*v_q^*)\ ,
\ee
and its associated Hamilton equations 
$\dot{v}_k=\{v_{k},\ov{\mathcal{A}}\}=-i\omega_{k}
\partial \ov{\mathcal{A}}/\partial v_{k}^{*}$ read
\be{vdotav}
\dot{v}_k=-i(\omega_k-k\Omega_1)v_k-i
\frac{\omega_k\mu}{4}\left(
\sum_{q=1}^{k-1}v_{k-q}v_q+
\sum_{q=1}^{N-k-1}2v_{k+q}v^*_q\right)\ ,
\ee
where, if $k=1$ or $k=N-1$, the first or the second sum on the right hand side are respectively absent.
Such equations must be solved, for $k=1,\dots,N-1$, with the initial condition 
\be{indatv}
v_{k}(0)=\delta_{k,1}\ .
\ee

Note that since the averaging \calleq{H3avform} is performed along the flow of the Hamiltonian $h_{\Omega_{1}}$, defined in 
\calleq{hOm1}, then $h_{\Omega_{1}}$ is a constant of motion of the averaged system:
$\{h_{\Omega_{1}},\ov{\mathcal{A}}\}=0$. Thus also
the quantity
\be{hom1}
h_{\omega_{1}}(v,v^{*})=\sum_{k=1}^{N-1}\left(
\frac{k\omega_{1}}{\omega_{k}}\right)|v_{k}|^{2}
\ee
is a constant of motion of the averaged system.

\subsection{Solution of the averaged system}
 
The averaged equations \calleq{vdotav} contain the small parameter $\mu$. We set
\be{Om1def}
\Omega_{1}(\mu)=\omega_{1}+\delta\omega_{1}(\mu)\ ,
\ee
the dependence on $\mu$ being left in the \emph{nonlinear
frequency shift} $\delta\omega_{1}=\Omega_{1}-\omega_{1}$.
We require that 
$\delta\omega_{1}\rightarrow0$ as $\mu\rightarrow0$, so that in absence of nonlinearity the initially excited mode will 
oscillate with its unperturbed frequency 
$\omega_{1}$. It follows\cite{N5} 
\be{triexpv}
v_k(t;\mu)=\mu^{k-1}\left(v^{(0)}_k(t)+\sum_{j\geq1}\mu^{2j}
v_k^{(j)}(t)\right)\ ,
\ee
\be{dom1exp}
\delta\omega_{1}(\mu)=\Omega_{1}(\mu)-\omega_{1}=
\sum_{j\geq1}\mu^{2j}\nu^{(j)}\ .
\ee
The initial condition $v_k(0)=\delta_{k,1}$ implies 
\be{incoeff}
v_k^{(j)}(0)=\delta_{k,1}\delta_{j,0}\ .
\ee
Upon substitution of \calleq{triexpv}-\calleq{dom1exp} in 
\calleq{vdotav} and balancing order by order 
one gets a nonlinear system of $N-1$ coupled equations
at any order $j\in\mathbb{N}$.
The first two orders ($j=0,1$) explicitly read:
\be{vk0}
\dot{v}_k^{(0)}=-i(\omega_k-k\omega_1)v_k^{(0)}-
i\frac{\omega_k}{4}
\sum_{q=1}^{k-1}v_{k-q}^{(0)}v_q^{(0)}\ ;
\ee
\bea
\dot{v}_k^{(1)}&=&-i(\omega_k-k\omega_1)v_k^{(1)}+
ik\nu^{(1)}v_{k}^{(0)} +\nonumber\\
&-&i\frac{\omega_k}{4}\left[
\sum_{q=1}^{k-1}\left(v_{k-q}^{(0)}v_q^{(1)}+
v_{k-q}^{(1)}v_q^{(0)}\right)
+2v_{k+1}^{(0)}\left(v_{1}^{(0)}\right)^{*}
\right]\ .\lab{vk1}
\eea
Again, a sum is absent if the upper limit of summation is
less than one. The latter two systems, in which the unknowns are
$v_{k}^{(0)}$, $v_k^{(1)}$ and $\nu^{(1)}$, have a triangular structure, i.e. the equations can be solved one after the other starting from the first down to the last one, with the initial condition 
\calleq{incoeff}. The expansion of the frequency shift
\calleq{dom1exp} (i.e. the coefficients $\nu^{(j)}$) is determined by requesting the vanishing of possible secular
terms in the expansion of $v_{1}$. One thus arrives at
the following leading order expansions:
\be{Om1sol}
\Omega_{1}(\mu)=\omega_{1}+
\frac{\mu^{2}\omega_{1}
\omega_{2}}{8(2\omega_{1}-\omega_{2})}+\dots;
\ee
\be{v1sol}
v_{1}(t)=1-
\frac{\mu^{2}\omega_{1}
\omega_{2}}{8(2\omega_{1}-\omega_{2})^{2}}
\left(1-e^{i(2\omega_{1}-\omega_{2})t}\right)+\dots;
\ee
\be{v2sol}
v_{2}(t)=\frac{\mu\omega_{2}}{4(2\omega_{1}-\omega_{2})}
\left(1-e^{i(2\omega_{1}-\omega_{2})t}\right)+\dots;
\ee
\be{v3sol}
v_{3}(t)=
\frac{\mu^{2}\omega_{2}\omega_{3}}{8(\omega_{2}-2\omega_{1})}
\left(
\frac{e^{i(3\omega_{1}-\omega_{3})t}-e^{i(2\omega_{1}-
\omega_{2})t}}{\omega_{3}-\omega_{2}-\omega_{1}}+
\frac{1-e^{i(3\omega_{1}-\omega_{3})t}}{\omega_{3}-3\omega_{1}}
\right)+\dots,
\ee
the dots standing for terms of higher order. 

\begin{figure}[!ht]
\centerline{\epsfig{figure=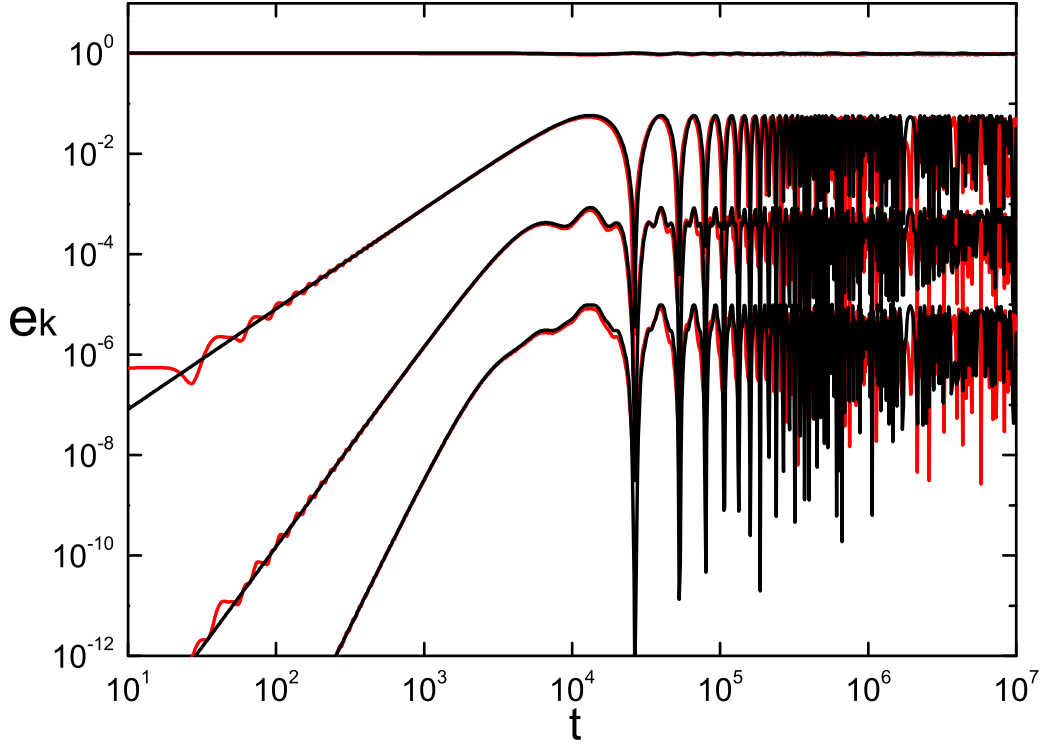,scale=0.6}}
\caption{$\en_k$ vs. $t$ (log-log scale), $k=1,2,3,4$,
for FPU, $E=10^{-4}$. Red curves: numerical data. Black
curves: analytic prediction.}
\label{fourmodes}
\end{figure}

In Fig.~\ref{fourmodes} the first four modal energies 
$\en_{k}(t)$ of the $\alpha$-model
are plotted vs. time $t$, in log-log scale, at the energy value $E=10^{-4}$. The red curves represent the numerical values,
while the four black lines correspond to the three analytic formulae \calleq{v1sol}-\calleq{v3sol} and to that of mode $k=4$, which is not explicitly reported here, since it has a very long, uninteresting expression. Note that 
$\en_{k}=|v_{k}|^{2}$. Note how the agreement is excellent up to the final computation time $t=10^{7}$. 

Fig.~\ref{fourmodes} shows that the perturbative
scheme used here works very well at low energies. 
The regime of validity of the perturbative solution can be estimated by requiring that in the expansions \calleq{Om1sol}
and \calleq{v1sol} of $\Omega_{1}$
and of the $v_{1}$, respectively, the leading order is dominant with respect to the correction proportional to $\mu^{2}$. The latter
condition turns out to be more restrictive than the former and,
by taking into account that
$2\omega_{1}-\omega_{2}=\pi^{3}/(4N^{3})+O(N^{-5})$, yields
\be{ENcond}
\alpha^{2}EN^{3}<\frac{\pi^{4}}{8}\simeq10\ ,
\ee
i.e. an upper bound to the energy which decreases with increasing 
$N$. For $N=32$ and $\alpha=0.33$, one gets 
$E<3\cdot10^{-3}$. Using \calleq{Om1sol}, condition 
\calleq{ENcond} can be transformed into 
\be{overlap}
\frac{\delta\omega_{1}}{2\omega_{1}-\omega_{2}}<\frac{1}{2}\ .
\ee
Thus the ratio of the nonlinear frequency shift 
$\delta\omega_{1}=\Omega_{1}-\omega_{1}$ to the 
\emph{resonance width} $2\omega_{1}-\omega_{2}$ must be
smaller than a constant, which is akin to the so-called resonance overlap criterion first formulated in Ref. \cite{IC} (notice that in the latter paper, dealing with the FPU-$\beta$ model, the considered resonance width was different; see also Refs. 
\cite{FP} and \cite{Shep} for comments on this point). 

For values of the energy larger than about $10^{-3}$ the agreement of the expressions \calleq{v1sol}-\calleq{v3sol} with the numerical data starts to worsen, and one would really need to include higher order corrections. However, some leading order analytic predictions concerning the behavior of the $v_{k}$'s can still be obtained, within the same perturbative scheme, either on very short times, or at saturation, as follows.

\subsection{Secular avalanche}

Since by the initial condition \calleq{incoeff} all the 
$v_k^{(0)}$ are zero at $t=0$ if $k\geq2$ and
$\omega_k-k\omega_1$ is small for low modes (and exactly vanishes for $k=1$), the contribution to the short term dynamics coming from the linear term $(\omega_k-k\omega_1)v_k^{(0)}$
on the right hand side of the equations \calleq{vk0} is expected to be small on short time. By neglecting such a term the leading order equations \calleq{vk0} simplify to
\be{saeqn}
\dot{v}_k^{(0)}=-i\frac{\omega_k}{4}
\sum_{q=1}^{k-1}v_{k-q}^{(0)}v_q^{(0)}\ ,
\ee
where again the sum on the right hand side is absent for $k=1$.
Equations \calleq{saeqn} with the initial condition $v_{k}^{(0)}=
\delta_{k,1}$ admit the exact (unique) solution
\be{vk0sa}
v_k^{(0)}(t)=(-it)^{k-1}c_k\ \ \ (k=1,\dots,N-1)\ ,
\ee
where the coefficients $c_k$ are computed by solving the nonlinear map 
\be{cmap}
c_1=1\ \ ;\ \ 
c_k=\frac{\omega_k}{4(k-1)}\sum_{q=1}^{k-1}c_{k-q}c_q
\ee
for $2\leq k\leq N-1$. Notice that to leading order, the first mode does not change from the initial condition: $v_1^{(0)}(t)=1$.
The evolution equation for the first mode at second order can be obtained by solving the second order equations \calleq{vk1} for $k=1$ with the \emph{choice} $\nu^{(1)}=0$.\cite{N6} Taking into account that $v_{1}^{(0)}=1$, 
and making use of \calleq{vk0sa} and \calleq{cmap}
to get $v_{2}^{(0)}=-itc_{2}$ and $c_{2}=\omega_{2}/4$,
one gets a differential equation for $v_{1}^{(1)}$, which can be integrated with the initial condition $v_{1}^{(1)}=0$. The result is
\be{v11sa}
v^{(1)}_{1}(t)=-\frac{\omega_{1}\omega_{2}t^{2}}{16}\ .
\ee
Finally, the normalized modal energies $\en_{k}=|v_{k}|^{2}$
can be computed:
\be{en1sa}
\en_1(t)=|v_{1}^{(0)}+\mu^{2}v_{1}^{(1)}+\dots|^{2}\simeq
1-\frac{\omega_1\omega_2(\mu t)^2}{8}\ ;
\ee
\be{enksa}
\en_k(t)=|\mu^{k-1}v_{k}^{(0)}+\dots|^{2}=
(\mu t)^{2(k-1)}c_k^2\ \ \ (k\geq2)\ ,
\ee
which explains the derivation of the law \calleq{e1t}-\calleq{ekt}
reported in Section II and tested in Fig.~\ref{ek_t001} of Section III.

\subsection{Role of resonance detuning and energy conservation}\label{roleres}

Let us discuss the validity
of the approximation obtained by neglecting the linear term on the right hand side
of equations \calleq{vk0}. 
This is done by substituting
on the right hand side of the same equation the solution 
\calleq{vk0sa} and by checking
whether and how long the linear term stays smaller than the quadratic term.
One thus finds that a first necessary condition for the approximation to be valid is that
\be{Tdetcond}
t<T^{det}_k\equiv \frac{k-1}{k\omega_1-\omega_k}\ \ \ (k\geq2)\ .
\ee
The 
\emph{detuning time} $T^{det}_k$ is the characteristic time within which
mode $k$ is resonantly pumped, the term proportional to the resonance amplitude
$k\omega_1-\omega_k$ being negligible. Now, $T^{det}_k$ is a monotonically
decreasing function of $k$, and for high mode numbers 
$T^{det}_k\approx N$, whereas for low modes $T^{det}_k\approx N^3/k^2$, which can be as large as $O(N^3)$ for the first few modes, actually too large to be compatible with the observed time-scale of formation of the quasi-state. When checking the validity of the approximation made by neglecting the linear term in \calleq{vk0}, one has to take into account that the quantity 
$h_{\omega_{1}}$, defined in \calleq{hom1}, is constant and equal to one. The latter condition implies that $|v_k|^2=\en_k<1$ for any $k\geq2$, yielding, when the (approximate) solution \calleq{vk0sa} is taken into account, a second necessary condition for the validity of the approximation made, namely
\be{Tsacond}
t<T^{sa}_k\equiv\frac{c_k^{-\frac{1}{k-1}}}{\mu}\ .
\ee
$T^{sa}_k$ is the mode dependent characteristic time 
of the secular avalanche process, i.e. the time within which mode $k$ can grow, when resonantly pumped
by the cascade process, without violating the conservation of the energy $h_{\omega_{1}}$.
We find that
$T^{sa}_k$ is a monotonically decreasing function of $k$ and,
by direct inspection of the map \calleq{cmap}, that 
$T^{sa}_k\sim 0.2N/\mu$, if $k$ is not too small. 
Note that from \calleq{en1sa} it follows that $\en_1>0$ if 
$t< T^{sa}_1\approx 2N/(\pi\mu)$, while there is obviously no detuning time for the first mode.

Having two characteristic times for each mode $k\geq2$, namely 
$T^{det}_k$ and $T^{sa}_k$, the actual characteristic time $T_k$ at which $\en_k$ stops its secular growth is defined as the minimum of the two for each $k$: 
$T_{k}\equiv\min\{T^{sa}_k,T^{det}_k\}$. For small values of $\mu$, $T_{k}=T^{sa}_k$ for $k<k_c$ and 
$T_{k}=T_k^{det}$ for $k>k_c$. An order of magnitude estimate of $k_c$ is obtained by observing that $T^{sa}_k\approx N/\mu$ for almost all modes, whereas 
$T^{det}_k\sim N^3/k^2$ if $k/N\ll 1$ (numerical factors are neglected). Then the crossover 
$T^{sa}_k=T^{det}_k$ takes place
at the critical mode number
\be{kc}
\frac{k_c}{N}\approx\sqrt{\mu}\ ,
\ee
valid if $k_c/N\ll 1$, i.e. if $\mu\ll 1$. The time-scale for the formation of the quasi-state is defined as the crossover time at $k_c$:
\be{Tqsdef}
T^{qs}\equiv T^{det}_{k_c}=T^{sa}_{k_c}\approx \frac{N}{\mu}
\ee
which is the estimate \calleq{Tqs}.

From the above reasonings, it follows that modes with 
$k<k_c$ are always inside a resonance layer and their modal energies stop to grow due to energy conservation, whereas modal energies $\en_k$ with $k>k_c$ stop to grow due to resonance detuning. We thus distinguish core modes with $k<k_c$ from those with $k>k_c$. Core modes share most of the energy in the quasi-state, and the resonant transfer of energy among them is effective. Higher modes outside the core are resonantly pumped for a certain time and are tuned out of resonance rather quickly. The quasi-state is characterized by
a few core modes sharing the energy and by an exponential localization of the m.e.s. for high modes. Notice that for $N=32$ and $\alpha=0.33$, equation \calleq{kc} gives $k_c$ varying from $2$ to $8$ as the energy varies from $0.01$ to $1$, in agreement 
with the number of modes that define the plateau of partial equipartition observable in the numerical spectra reported in the left column of Fig.~\ref{multirun}.

\subsection{Fixed point of the averaged system: quasi-state}

Let us consider the averaged equations \calleq{vdotav}. 
In the limit $\mu\rightarrow0$ one has $\ov{\mathcal{H}}_{3}
\rightarrow0$, we require $\Omega_{1}\rightarrow\omega_{1}$, and the equations simplify to
\be{vdotav0}
\dot{v}_{k}=-i(\omega_{k}-k\omega_{1})v_{k}\ ,
\ee
whose solution with initial condition \calleq{indatv} is
$v_{k}(t)=v_{k}(0)=\delta_{k,1}$. In other words, the system is placed at $t=0$ on a fixed point and stays there forever. Notice that system \calleq{vdotav0} admits a one-parameter family of fixed points
\be{C0}
C_{0}=\{|v_{1}|=1\ ;\ |v_{k}|=0\ ,\ k\geq2\}\ ,
\ee
which is a unit circle in the $v_{1}$-plane of $\mathbb{C}^{N-1}$.
It can be easily checked that the set $C_{0}$ is the maximum of the Hamiltonian $\mathcal{H}_{2}-h_{\omega_{1}}$
(limit of the Hamiltonian $\ov{\mathcal{A}}$ as $\mu\rightarrow0$)
constrained to the set $h_{\omega_{1}}=1$. 
If $\mu>0$ but small enough, the 
set $C_{0}$ is slightly deformed to the set $C_{\mu}$
of stationary points of the full averaged system \calleq{vdotav}
on the set $h_{\omega_{1}}=1$, i.e. the set of critical points 
of the Hamiltonian $\ov{\mathcal{A}}=\mathcal{H}_{2}-
h_{\Omega_{1}}+\ov{\mathcal{H}}_{3}$ constrained to the set
$h_{\omega_{1}}=1$. Moreover, if $\mu$ is small enough, the critical set $C_{\mu}$ preserves its maximum property (i.e. 
$C_{\mu}$ is the maximum of $\ov{\mathcal{A}}$ on 
$h_{\omega_{1}}=1$) and turns out to be Lyapunov stable: if 
$v$ is initially close to $C_{\mu}$ it stays close to it forever. On the other hand, the initial condition $v_{k}(0)=\delta_{k,1}$ means that one starts on $C_{0}$, so that if $C_{0}$ and $C_{\mu}$ are close, i.e. if $\mu$ is small enough, the orbit of the averaged system stays forever close to $C_{\mu}$. We make use of this qualitative
(but exact) reasoning to identify the quasi-state orbit of the 
$\alpha$-model with the set $C_{\mu}$. 

The fixed points of system \calleq{vdotav} 
can be perturbatively computed by means of 
the expansions \calleq{triexpv}-\calleq{dom1exp}. To leading order, one has to
find the fixed point of system \calleq{vk0}, setting 
$\dot{v}_{k}^{(0)}$, which yields 
\be{qseqn0}
v_{1}^{(0)}=\lambda\ \ ;\ \ v_{k}^{(0)}=\frac{\omega_{k}}{4(k\omega_{1}-\omega_{k})}\sum_{q=1}^{k-1}v_{k-q}^{(0)}v_q^{(0)}\ ,
\ (k\geq2)
\ee
where $\lambda$, at this stage, is an arbitrary complex number
such that $|\lambda|^{2}=\en_{1}$, to leading order.
Isolating the dependence of $v_{k}^{(0)}$ on $\lambda$ one gets
\be{vk0lam}
v_{k}^{(0)}=\lambda^{k}f_{k}\ ,
\ee
where the real coefficients $f_k$ are found by solving the nonlinear map problem
\be{fmap}
f_1=1\ \ ;\ \ 
f_k=\frac{\omega_k}{4(k\omega_1-\omega_k)}
\sum_{q=1}^{k-1}f_{k-q}f_q
\ee
for $2\leq k\leq N-1$. Recalling that, to leading order, 
$v_{k}=\mu^{k-1}v_{k}^{(0)}$, one can compute the m.e.s.
\be{spec}
\en_k=|v_{k}|^{2}=
f_k^2\left(\mu^2\en_1\right)^{k-1}\en_1\ .
\ee
Such a spectrum depends on the unknown 
$\en_{1}=|\lambda|^{2}$, which is determined by the 
normalization condition $h_{\omega_{1}}=1$, i.e.
\be{normh2}
h_{\omega_{1}}=\sum_{k=1}^{N-1}\left(\frac{k\omega_1}{\omega_k}\right)
\en_k=
\sum_{k=1}^{N-1}\left(\frac{k\omega_1}{\omega_k}\right)
f_k^2\left(\mu^2\en_1\right)^{k-1}\en_1=1\ .
\ee
The latter condition determines the modulus of $\lambda$,
so that an overall phase remains undetermined. For this reason, the set $C_{\mu}$ is composed by a one-parameter family
of stationary points of the averaged system.

We now observe that to a stationary solution of the averaged system, there corresponds a periodic orbit, or one-dimensional torus, of the original system:
recall that, before averaging, we performed the change of variables $u_{k}=e^{-ik\Omega_{1}t}v_{k}$. The fundamental frequency $\Omega_{1}$ of the quasi-state periodic orbit is easily
determined by means of equations \calleq{vk1}. Setting there
$k=1$ and $\dot{v}_{1}^{(1)}=0$, and 
making use of \calleq{vk0lam}-\calleq{fmap}, one gets 
\be{v1qs}
\nu^{(1)}=\frac{\omega_{1}v_{2}^{(0)}(v_{1}^{(0)})^{*}}{2
v_{1}^{(0)}}=
\frac{\en_{1}\omega_{1}\omega_{2}}{8(2\omega_{1}-\omega_{2})}
\ ,
\ee
which implies 
\be{Om1qs}
\Omega_{1}=\omega_{1}+\mu^{2}\nu^{(1)}=
\omega_{1}+\frac{\mu^{2}\en_{1}\omega_{1}\omega_{2}}{
8(2\omega_{1}-\omega_{2})}\ .
\ee
Notice that in the regime of strong localization, when 
$\en_{1}\simeq1$, the above expression for the corrected frequency coincides with that given in \calleq{Om1sol}.

As already mentioned, the quasi-state m.e.s. \calleq{spec}
fits the numerically obtained  one at low energy, when the exponential localization is so strong that only a few low modes contribute to
the total energy. In this case, the map \calleq{fmap}
can be approximately solved, as follows. First of all,
notice that if only low (acoustic) modes
are relevant, one can substitute the approximate expressions
$\omega_{k}\simeq\pi k/N$, $k\omega_{1}-\omega_{k}\simeq
\pi^{3}k(k^{2}-1)/(24N^{3})$
in the numerator and denominator of \calleq{fmap}, respectively. Second, we look for a solution of the map in the form
$f_{k}=k\ g^{k-1}$, which fits the condition $f_{1}=1$. Upon substitution of the latter expression into 
\calleq{fmap}, and after some simple calculation, one gets
$g=\pi^{2}/N^{2}$. Thus, in the regime of strong localization, the approximate solution of the map \calleq{fmap} is given by 
$f_k\simeq k(N^2/\pi^2)^{k-1}$. As a consequence, 
recalling that $\mu^{2}=\alpha^{2}E/N$ and that 
$\en_{1}=E_{1}/E$, for the spectrum \calleq{spec} one gets the approximate expression
\be{specqb}
\en_{k}=k^{2}\left(\frac{\alpha^{2}EN^{3}\en_{1}}{\pi^{4}}
\right)^{k-1}\en_{1}\ .
\ee
Under the same acoustic approximation ($\omega_k\simeq k\omega_1$) leading to the latter formula, condition
\calleq{normh2} becomes
\be{normhqb}
\sum_{k=1}^{N-1}\en_{k}=\sum_{k=1}^{N-1}
k^{2}\left(\frac{\alpha^{2}EN^{3}\en_{1}}{\pi^{4}}\right)^{k-1}
\en_{1}=1\ .
\ee
This explains formulas \calleq{qb1}-\calleq{Om1} given in Section
II. We stress again that the m.e.s. \calleq{specqb} corresponds to the leading order spectrum of the Lyapunov continuation to the nonlinear case of the first mode, first reported in Ref. 
\cite{FIK1}
and there named $q$-breather. 
We also recall here that the normalization condition 
\calleq{normhqb} can be simply solved. Indeed, by defining 
$x\equiv\alpha^2EN^3/\pi^4$ and 
$y\equiv\alpha^2E_1N^3/\pi^4$, it is easily found that 
\calleq{normhqb}, in the limit of large $N$ and for $y<1$, yields the simple implicit equation $y(1+y)/(1-y)^3=x$.
This is explicitly solved, for $x\geq0$ (and $0\leq y<1$), to give
$y=F(x)$, for a suitable function $F$,
which determines $E_1$ (and thus $\en_1$) in terms of the total energy $E$, of $\alpha$ and of the number of particles $N$. In particular, having determined $F(x)$, it turns out that, at any fixed $x$, $\en_1=F(x)/x$,
which can then be used to finally determine the m.e.s. 
\calleq{specqb}.

To conclude this subsection, we observe that all the above construction was performed to leading order, and applied to
the Toda model, would give the same result. This is a consequence of the coincidence to third order of the FPU-$\alpha$
and the Toda models expressed by relation \calleq{HaHT}.

\subsection{Long term stability of the quasi-state}

The argument used to guess that the one parameter family 
$C_{\mu}$ of stationary points of the Hamiltonian \calleq{Aav} constrained to the set
$h_{\omega_{1}}=1$ is the candidate quasi-state orbit of the 
FPU-$\alpha$ and Toda models, is a qualitative one. We did not specify how small the parameter $\mu$ has to be to ensure that $C_{\mu}$ is a maximum of the constrained averaged Hamiltonian and moreover the unperturbed orbit $C_{0}$ (which contains the initial datum) is in the domain of concavity 
of $C_{\mu}$. In this respect, we do not provide any precise estimate on $\mu$, which would require technical arguments that are beyond the aim of the present paper. We only mention that, from the discussion made in Section IV C above, one may expect that a condition of the form $E=O(N^{-3})$
should hold to ensure the Lyapunov stability of $C_{\mu}$.

Apart from the precise energy-size dependence of the stability domain of the one-dimensional torus $C_{\mu}$, it is important to understand that the construction made above is a leading order one, which ensures, for the real system, confinement close to the set $C_{\mu}$ on a relatively short time-scale, typically an inverse power law of the small parameter $\mu$. In order to explain the 
stability on much longer times observed numerically, one should go beyond the leading order construction, which modifies the averaged Hamiltonian \calleq{Aav} with the addition of higher order terms, and increases the length of the confinement time.
Such an approach was used in Refs. \cite{BN,Nek1,Nek2}
to prove that in the case of initial excitation of the first mode in a suitable class of string equations, the confinement time
grows as the exponential of the inverse of the small parameter.
However, it must be stressed that importing those techniques
to the problem at hand is nontrivial at all, essentially due to the lack of complete resonance in the FPU problem (explicit reference to the FPU problem, in this perspective, has been made in Ref. \cite{Nek1}). 

\section{Concluding remarks}

We have shown that the classical FPU paradox
can be easily understood if one compares the behavior of 
the (presumably) nonintegrable FPU-$\alpha$-model with that of the integrable Toda model, for the same initial condition, at the same energy. One thus finds that the dynamics of the $\alpha$-model on a short initial time interval is very close to that of the Toda model. The quasi-state guessed by FPU is nothing but the quasi-periodic Toda torus on which the dynamics is initiated for the same initial condition. At very low energy, such a torus becomes one-dimensional and the FPU-$\alpha$ system evolves close to it up to extremely long times. As the energy increases, the dimension of the reference torus grows, while the time of confinement close to it (or stability time) for the FPU-$\alpha$ system
decreases. Finally, the approach to a state of energy equipartition of the FPU-$\alpha$ model takes place through a diffusive-like detachment from the quasi-state, a phenomenon that
is  chaotic in character, and consequently, absent in the integrable model. 

Regarding the FPU $\alpha$-model as a perturbation of the Toda one, as first suggested in Ref. \cite{FFML}, and strongly motivated here, seems to be quite promising. For example, in 
Ref. \cite{HK09} it is shown that the Toda Hamiltonian fits the hypotheses requested to the unperturbed Hamiltonian in the Nekhoroshev theorem, with the purpose to explain the phenomenology of the FPU-$\alpha$ model in terms of the exponential stability of all the Toda integrals (or actions). However, precise/optimal estimates of the regime of validity (energy-size domain) of such a conclusion are still lacking. Of course, if the energy is small enough, one cannot exclude the possibility to enter a KAM regime, i.e. to have stability over infinitely long times. Here also, we know that this is possible in principle\cite{Rink}, but again, precise/optimal estimates of the regime of validity are not available yet. We finally notice that it is not presently clear whether this approach (i.e. considering the FPU system as a perturbation of the Toda one) is valid for initial conditions involving also high frequency modes or even for generic excitations.

One interesting question is whether and how the energy region for the long-term stability of the quasi-state collapses or not to zero by increasing $N$. In the light of the purely numerical results contained in Ref. \cite{BP11}, a first answer seems to be that exponentially long times to equipartition might be, for the $\alpha$-model, a finite-size effect, observable only in an unspecified regime $E\rightarrow0$ as $N\rightarrow\infty$. In the same paper, it is shown that for particular low frequency initial data, the $\beta$-model seems to exhibit exponentially long times to equipartition in the thermodynamic limit. On the other hand, in Ref. \cite{CM11}, an analytic lower bound to the canonical mixing time of a weakly coupled $\phi^{4}$ model is provided, showing that such time is longer than a stretched exponential of the inverse of the temperature. The full scenario is quite complicated and deserves a lot of further study. 

\paragraph{acknowledgments}
This work was initiated and in large part completed during a stay of A.P. and H.C. in the MPIPKS of Dresden, whose facilities were used to perform all the numerical calculations. The authors thank 
D. Bambusi, G. Benettin, A. Giorgilli and T. Penati for useful discussions on the subject, and are also indebted to D. Ryabov
for important remarks on the Toda model.


\begin{thebibliography}{70}



\bibitem{FPU} E. Fermi, J. Pasta and S. Ulam, LA-1940 internal report, 1955.
Reprinted in the \emph{Collect Papers of E. Fermi}, Vol. II, University of Chicago Press and Accademia Nazionale dei Lincei, 1965, 978-988.

\bibitem{Fetal} E. Fucito, F. Marchesoni, E. Marinari, G. Parisi, L. Peliti, S. Ruffo and A. Vulpiani, J. de Phys. {\bf 43},
707-713 (1982).

\bibitem{LPRVms} R. Livi, M. Pettini, S. Ruffo and A. Vulpiani,
Phys. Rev. A {\bf 31}, 2740-2742 (1985).

\bibitem{BGG} L. Berchialla, L. Galgani and A. Giorgilli, DCDS-A
{\bf 11}, 855-866 (2004).

\bibitem{FIK1} S. Flach, M. V. Ivanchenko and O. I. Kanakov, Phys. Rev. Lett. {\bf 95}, 064102/1-4 (2005).

\bibitem{FIK2} S. Flach, M. V. Ivanchenko and O. I. Kanakov, Phys. Rev. E {\bf 73}, 036618/1-14 (2006).

\bibitem{FKIM} S. Flach, O. I. Kanakov, M. V. Ivanchenko and K. G. Mishagin, Int. J. Mod. Phys. B {\bf 21}, 3925-3932 (2007).
 
\bibitem{PF} T. Penati and S. Flach, Chaos {\bf 17}, 
023102/1-16 (2007).

\bibitem{FP} S. Flach and A. Ponno, Physica D {\bf 237}, 
908-917 (2008).

\bibitem{ZK} N. J. Zabusky and M. D. Kruskal, Phys. Rev. Lett. {\bf 15}, 240-243 (1965).

\bibitem{Shep} D. L. Shepelyansky, Nonlinearity {\bf 10}, 1331-1338 (1997).

\bibitem{P03} A. Ponno, Europhys. Lett. {\bf 64}, 606-612
 (2003).

\bibitem{P05} A. Ponno, in \emph{Chaotic Dynamics and Transport in Classical and Quantum Systems}, 
P. Collet et al. (eds.); Kluwer Academic Publishers,
431-440 (2005).

\bibitem{BPcmp} D. Bambusi and A. Ponno, Commun. Math. Phys. {\bf 264}, 539-561 (2006).

\bibitem{BLP} G. Benettin, R. Livi and A. Ponno, J. Stat. Phys.
{\bf 135}, 873-893 (2009).

\bibitem{FFML} W. E. Ferguson Jr., H. Flaschka and D. W. McLaughlin, J. Comp. Phys. {\bf 45}, 157-209 (1982).

\bibitem{ILRV} S. Isola, R. Livi, S. Ruffo and A. Vulpiani, Phys. Rev. A {\bf 33}, 1163-1170 (1986).

\bibitem{CCPC} L. Casetti, M. Cerruti-Sola, M. Pettini and
E. G. D. Cohen, Phys. Rev. E {\bf 55}, 6566-6574 (1997).

\bibitem{GPP} A. Giorgilli, S. Paleari and T. Penati, DCDS-B
{\bf 5}, 991-1004 (2005).

\bibitem{Zab06} N. J. Zabusky, Z. Sun and G. Peng, Chaos
{\bf 16}, 013130/1-12 (2006).

\bibitem{TodaPR} M. Toda, Phys. Rep. {\bf 18}, 1-124 (1975).

\bibitem{FST} J. Ford, S. D. Stoddard and J. S. Turner,
Prog. Theor. Phys. {\bf 50}, 1547-1560 (1973).

\bibitem{He} M. H\'{e}non, Phys. Rev. B {\bf9}, 1921-1923 (1974).

\bibitem{Fl} H. Flaschka, Phys. Rev. B {\bf9}, 1924-1925 (1974).

\bibitem{PL} M. Pettini and M. Landolfi, Phys. Rev. A 
{\bf 41}, 768-783 (1990).

\bibitem{DLR} J. De Luca, A. J. Lichtenberg and S. Ruffo,
Phys. Rev. E {\bf 60}, 3781-3786 (1999).

\bibitem{BGP} L. Berchialla, A. Giorgilli and S. Paleari, Phys. Lett. A {\bf 321}, 167-172 (2004).

\bibitem{BP11} G. Benettin and A. Ponno, J. Stat. Phys. 
{\bf 144}, 793-812 (2011).

\bibitem{FordPR} J. Ford, Phys. Rep. {\bf 213}, 271-310 (1992).

\bibitem{CHAOS} Focus Issue \emph{The ``Fermi-Pasta-Ulam''-problem: the first 50 years}, Chaos {\bf 15} (2005).

\bibitem{LNP} \emph{The Fermi-Pasta-Ulam Problem: A Status Report}, Lect. Notes Phys. {\bf 728}, G. Gallavotti ed.; Springer-Verlag, 2008.

\bibitem{Yo1} H. Yoshida, Phys. Lett. A {\bf 150}, 262-268 (1990).

\bibitem{Yo2} H. Yoshida, Cel. Mech. Dyn. Astron. {\bf 56}, 
27-43 (1993).

\bibitem{N0} The time-step $\tau$ must be much smaller than $\pi$, the latter being the minimum normal mode period (the maximum frequency is $2$). The fourth order algorithm ensures energy conservation with an absolute error of the order
$\tau^4E^2$. For details on the numerical integration of FPU
systems see e.g.: G. Benettin and A. Ponno, Physica D {\bf 240}, 568-573 (2011).

\bibitem{N1} The value of $\alpha$ is irrelevant, since it can be set to one by the canonical transformation $(q,p,H,t)\mapsto(\alpha q,\alpha p,\alpha^2 H,t)$, which also shows that the actual parameter of nonlinearity in the problem is
the product $\alpha^2E$.

\bibitem{N2} Formula \calleq{ET} can be obtained by first computing $q_{n+1}-q_{n}$ for the initial condition 
\calleq{incond}, expanding the exponential in the Toda potential 
\calleq{HT}, and then summing first over $n$.

\bibitem{AbSt} M. Abramowitz and I. A. Stegun, \emph{Handbook of Mathematical Functions}, Dover, 1965.

\bibitem{N3} In the regime explored in the present paper $A\simeq A'$, or, which is the same, $E_\alpha(A)\simeq E_T(A)$. Indeed, since $I_{0}(x)=1+x^{2}/4+x^{4}/64+\dots$, from \calleq{E} and 
\calleq{ET} it follows that
$E_{T}(A)=E_{\alpha}(A)[1+\alpha^2E_{\alpha}(A)/N+\dots]$, and the quantity $\alpha^2E_{\alpha}(A)/N$ is always small in our runs,
being at most $10^{-2}$ down to $10^{-5}$, and even lower in some cases.

\bibitem{N4} Since the cubic potential is not lower bounded, the micro-canonical measure of the $\alpha$-model does not exist and the system breaks-down in a finite time when a sufficient amount of energy localizes on a few particles. As a matter of fact, in the quasi-linear regime considered here, the break-down takes place after
equipartition has been reached, the latter state being well described by the micro-canonical measure of the harmonic chain. In all the numerical runs reported in the sequel the break-down was never reached.

\bibitem{Ford61} J. Ford, J. Math. Phys. {\bf 2}, 387-393 (1961).

\bibitem{LPRV} R. Livi, S. Ruffo, M. Pettini and A. Vulpiani, Il Nuovo Cimento B {\bf 89}, 120-130 (1985).

\bibitem{Zab62} N. J. Zabusky, J. Math. Phys. {\bf 3}, 1028-1039 (1962).

\bibitem{KZ64} M. D. Kruskal and N. J. Zabusky, J. Math. Phys. {\bf 5}, 231-244 (1964).
 
\bibitem{CEB1} H. Christodoulidi, C. Efthymiopoulos and T. Bountis, Phys. Rev. E {\bf 81}, 016210/1-16 (2010).

\bibitem{CEB2} H. Christodoulidi, C. Efthymiopoulos and T. Bountis; preprint (2011).

\bibitem{BM} N. N. Bogolyubov and Y. A. Mitropolsky,
\emph{Asymptotic Methods in the Theory of Non-linear Oscillations}, Gordon \& Breach, NY, 1961.

\bibitem{SVM} J. A. Sanders, F. Verhulst and J. Murdock, 
\emph{Averaging Methods in Nonlinear Dynamical
Systems}, Springer, 2007.

\bibitem{N5} As can be proved by induction, 
\calleq{triexpv} and \calleq{dom1exp} are the only possible formal power series expansions compatible with the equations \calleq{vdotav} and with the initial condition \calleq{indatv}.

\bibitem{IC} F.M. Izrailev and B.V. Chirikov,
Sov. Phys. Dokl. {\bf 11}, 30-34 (1966).

\bibitem{N6} The quantity $\nu^{(1)}$ is an unknown of the problem and there is more than one way to determine its value; by setting the latter value to zero means giving up to control secular divergencies, which still yields an approximate solution, though valid over very short times.

\bibitem{BN} D. Bambusi and N. N. Nekhoroshev, Physica D 
{\bf 122}, 73-104 (1998).

\bibitem{Nek1} N. N. Nekhoroshev, in the Proceedings of the Porquerolles School
2001 on \emph{Hamiltonian Systems and Fourier Analysis}, ed. by D. Benest
et al., Cambridge Scientific Publishers, 289-302 (2005).

\bibitem{Nek2} N. N. Nekhoroshev, Trans. Moscow Math. Soc. of the year 2002, 151-217.

\bibitem{HK09} A. Henrici and T. Kappeler, Chaos {\bf 19},
033120/1-13 (2009).

\bibitem{Rink} B. Rink, Commun. Math. Phys. {\bf 261}, 
613-627 (2006).

\bibitem{CM11} A. Carati and A. Maiocchi; preprint (2010)
on {\tt http://arxiv.org/abs/1011.5846}

\end{thebibliography}
\end{document}